\documentclass[letterpaper]{article} % DO NOT CHANGE THIS
\usepackage{Style/arxiv}  % DO NOT CHANGE THIS
\usepackage{times}  % DO NOT CHANGE THIS
\usepackage{helvet}  % DO NOT CHANGE THIS
\usepackage{courier}  % DO NOT CHANGE THIS
\usepackage[hyphens]{url}  % DO NOT CHANGE THIS
\usepackage{graphicx} % DO NOT CHANGE THIS
\usepackage{natbib}  % DO NOT CHANGE THIS AND DO NOT ADD ANY OPTIONS TO IT
\usepackage{caption} % DO NOT CHANGE THIS AND DO NOT ADD ANY OPTIONS TO IT
\usepackage{algorithm}
\usepackage{algorithmic}

\usepackage{newfloat}
\usepackage{listings}
\DeclareCaptionStyle{ruled}{labelfont=normalfont,labelsep=colon,strut=off} % DO NOT CHANGE THIS
\floatstyle{ruled}
\newfloat{listing}{tb}{lst}{}
\floatname{listing}{Listing}
\newcommand{\arxiv}[0]{}
\usepackage{xcolor}
\usepackage{colortbl}
\usepackage[hidelinks]{hyperref}
\usepackage{booktabs}
\usepackage{amsmath}
\usepackage{amssymb}
\usepackage{bm}
\usepackage{subcaption}
\title{Quality-Diversity Generative Sampling\\
for Learning with Synthetic Data}
\author {
    Allen Chang\textsuperscript{\rm 1},
    Matthew C. Fontaine\textsuperscript{\rm 1},
    Serena Booth\textsuperscript{\rm 2},
    Maja J. Matari\'c\textsuperscript{\rm 1},
    Stefanos Nikolaidis\textsuperscript{\rm 1}
}
\affiliations {
    \textsuperscript{\rm 1} University of Southern California, Los Angeles, USA\\
    \textsuperscript{\rm 2} Massachusetts Institute of Technology, Cambridge, USA\\
    \ifdefined\arxiv{
    \texttt{\{changall, mfontain, mataric, nikolaid\}@usc.edu},
    \texttt{sbooth@mit.edu}
    }\else{
    \{changall, mfontain, mataric, nikolaid\}@usc.edu,
    sbooth@mit.edu
    }\fi
}

\begin{document}

\maketitle

\begin{abstract}
Generative models can serve as surrogates for some real data sources by creating synthetic training datasets, but in doing so they may transfer biases to downstream tasks.
We focus on protecting quality and diversity when generating synthetic training datasets.
We propose quality-diversity generative sampling (QDGS), a framework for sampling data uniformly across a user-defined measure space, despite the data coming from a biased generator.
QDGS is a model-agnostic framework that uses prompt guidance to optimize a quality objective across measures of diversity for synthetically generated data, without fine-tuning the generative model.
Using balanced synthetic datasets generated by QDGS, we first debias classifiers trained on color-biased shape datasets as a proof-of-concept. 
By applying QDGS to facial data synthesis, we prompt for desired semantic concepts, such as skin tone and age, to create an intersectional dataset with a combined blend of visual features.
Leveraging this balanced data for training classifiers improves fairness while maintaining accuracy on facial recognition benchmarks.
Code available at: 
\ifdefined\arxiv{\href{https://github.com/Cylumn/qd-generative-sampling}{\texttt{github.com/Cylumn/qd-generative-sampling}}}\else{https://github.com/Cylumn/qd-generative-sampling}\fi.
\end{abstract}

\section{Introduction}
\begin{figure}[tbp]
\centering
\includegraphics[width=\columnwidth]{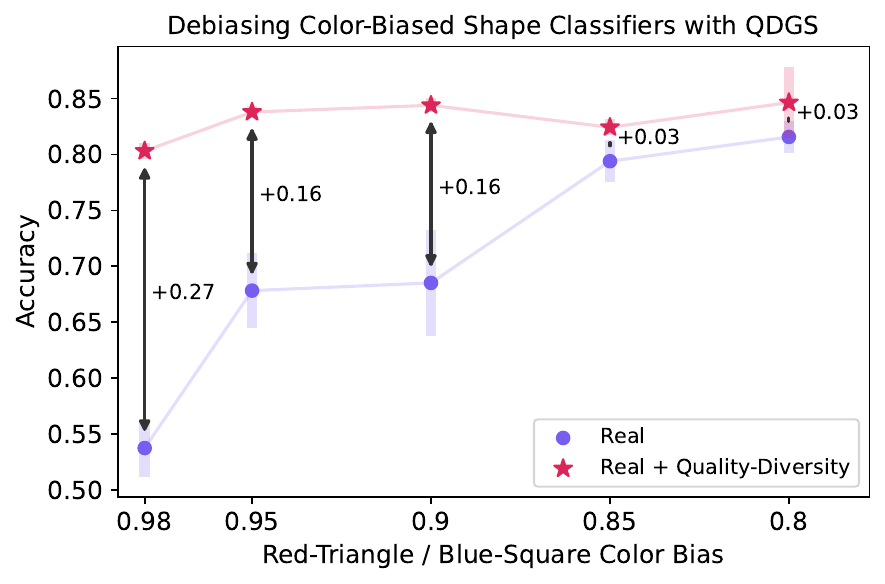}
\caption{Fine-tuning color-biased shape classifiers with balanced data generated from QDGS results in higher accuracy across minority shape-color combinations. Standard error bars across $5$ trials are shaded in.}
\label{fig:cover}
\end{figure}

Generative models can produce large amounts of synthetic data at low cost, making them viable surrogates for some real data sources in model training~\cite{google-imagenet,phillip-stable}.
However, generative models suffer from representational biases~\cite{dalle-bias, fair-diffusion,mode-collapse}, endangering a transfer of bias to downstream tasks~\cite{bias-exacerbate}.
We propose a sampling framework that protects quality and diversity when generating synthetic datasets, and we show that these datasets can suppress performance decline on minority groups (Fig.~\ref{fig:cover}).

Balancing synthetic datasets is not a trivial task, since generative models harbor biases that relate to different semantic concepts.
Some biases result from training deep generative models to replicate Internet data, where there are common issues of sampling selection bias.
Internet data also contain social biases that result from disparities in language use among social groups, eliciting stereotypes and prejudices~\cite{social-bias}.
Biases are intensified by selecting training approaches that amplify class and feature imbalances, such as some data augmentation and regularization methods~\cite{lecun-bias}.
Thus, most existing generative models suffer from multiple biases that harm the creation of fair synthetic datasets.
When creating these datasets for training downstream models, it is helpful to simultaneously consider multiple of these biases.

\begin{figure*}[tbp]
\centering
\includegraphics[width=\textwidth]{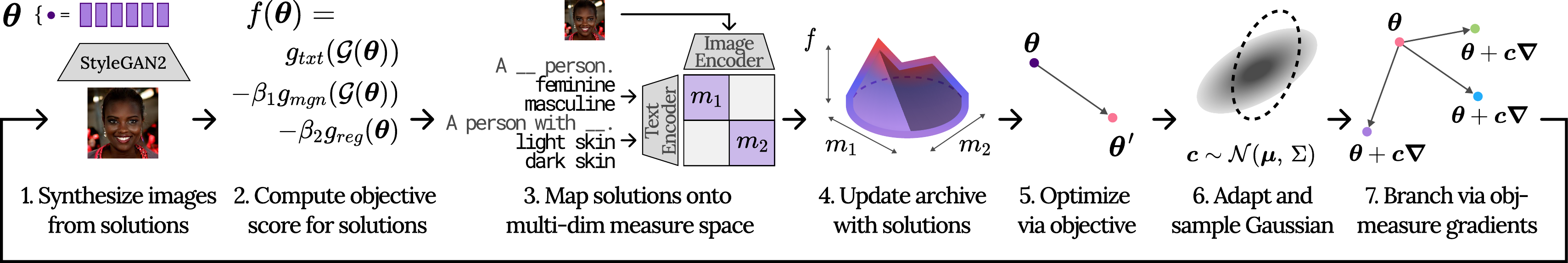}
\caption{An implementation of the data generation pipeline used in QDGS. Steps 1-4 describe how solutions from QDGS are mapped to the objective-measure archive, and steps 5-7 describe how solutions are updated.}
\label{fig:overview}
\end{figure*}

However, prior methods for sampling debiased synthetic data tend to focus on removing a single bias at a time, through single-factor diversity optimization~\cite{mcduff-bayesian, jain-pretrain}.
By increasing diversity along one dimension at a time, sampling methods miss the opportunity to generate intersectional datasets.

We leverage the insight that a generative model's high dimensionality latent space can often combine underrepresented semantic concepts to create synthetic datasets with a rich blend of intersectional data.
However, the same latent spaces typically collapse on modes that disproportionately represent the dominant classes or data attributes.
To address these biases, we propose exploring the latent space to identify and generate underrepresented attribute combinations.
We call this approach {\it quality-diversity generative sampling (QDGS)}, a framework using quality-diversity optimization with prompt guidance for producing synthetic datasets with a balanced spread of desired features.

To better understand how to improve parity across different semantic concepts, we consider the domain of facial images. 
Randomly sampling from StyleGAN2~\cite{stylegan} reproduces a skin tone imbalance ($\sim\!7\!:\!1$ light\footnote{Classifiers trained on Fitzpatrick Skin Type annotations from the IJB-C~\cite{ijb} are used to recognize skin tone. Images recognized as types 1 or 2 are denoted as \textit{light}, types 3 or 4 are denoted as \textit{mixed}, and types 5 or 6 are denoted as \textit{dark}.} to dark skin tone ratio) and an age imbalance ($\sim\!2\!:\!1$ young to old ratio) from its training dataset Flickr-Faces-HQ (FFHQ)~\cite{ffhq}.
Synthetic facial datasets typically contain other demographic and visual disparities (Table~\ref{tab:datasets-and-sampling}), which can create multiple dimensions of bias.

By employing user-defined language prompts to describe diversity measures (e.g., ``\textit{dark skin tone},'' ``\textit{smiling},'', ``\textit{dim lighting},'' etc.) and an objective (e.g., ``\textit{detailed image}''), QDGS enables a greater degree of control for creating synthetic training data.
We express each measure as an independent dimension so that coverage extends to intersectional groups (e.g., ``\textit{feminine}'' and ``\textit{dark skin tone}'' and ``\textit{short hair}'').
When applied to StyleGAN2, QDGS increases the proportion of images recognized with dark skin tones from $9.4\%$ to $25.2\%$, images recognized as old from $37.0\%$ to $45.7\%$, and achieves a more uniform spread of gender\footnote{We use the term \textit{gender} to refer to a 1D spectrum of facial presentation styles. We denote the ends of the spectrum as \textit{masculine} and \textit{feminine}, and we refer to the center as \textit{androgynous}. Gender attributes, among other continuous demographic attributes, are sometimes discretized into bins for analysis.} presentations and hair length.
Notably, QDGS is model-agnostic and adaptable, allowing for its application to various tasks.
This is achieved by interchanging the generative model and language prompts according to the desired attributes suited for the training data of the task at hand.

QDGS has the potential to improve trained classifiers.
As a proof-of-concept, we show that QDGS can debias an imbalanced dataset of colors and shapes, resulting in up to $\approx\!27\%$ improvement on downstream classification accuracy.
In the more challenging setting of facial recognition, where models typically perform better for light-skinned users, 
QDGS is able to increase accuracy on dark-skinned faces from $88.08\%$ to $88.94\%$.
Finally, QDGS achieves the highest average accuracy across facial recognition benchmarks for models trained on biased datasets.
These results support training with balanced synthetic datasets.

\begin{table*}[tbp]
  \centering
  \setlength{\tabcolsep}{2pt}
  \begin{tabular}{l ccc c cc c ccc c cc c ccc}
    \toprule
    \textbf{Dataset} &
    \multicolumn{3}{c}{\textbf{Skin Tone}} &&
    \multicolumn{2}{c}{\textbf{Age}} &&
    \multicolumn{3}{c}{\textbf{Gender}} &&
    \multicolumn{2}{c}{\textbf{Background}} &&
    \multicolumn{3}{c}{\textbf{Face Rotation}}
    \\
    &
    \textit{\small{Light}} & \textit{\small{Mixed}} & \textit{\small{Dark}} & \hspace{2pt} & 
    \textit{\small{Young}} & \textit{\small{Old}} & 
    \hspace{2pt} &
    \textit{\small{Masc}} & \textit{\small{Andro}} & \textit{\small{Fem}} & \hspace{2pt} & 
    \textit{\small{Indoors}} & \textit{\small{Outdoors}} & 
    \hspace{2pt} &
    \textit{\small{Frontal}} & \textit{\small{Turned}} & 
    \textit{\small{Profile}}
    \\
    \toprule
    \small{Facial Benchmarks (Large Scale)}\\
    \cmidrule(r){1-1}
LFW & $72.7$ & $18.8$ & $08.5$ &  & $29.6$ & $70.4$ &  & $75.0$ & $01.0$ & $24.0$ &  & $60.2$ & $39.9$ &  & $74.2$ & $23.9$ & $01.9$\\
CelebA & $71.5$ & $20.0$ & $08.5$ &  & $70.9$ & $29.1$ &  & $41.4$ & $00.5$ & $58.2$ &  & $\pmb{49.9}$ & $\pmb{50.1}$ &  & $53.6$ & $41.2$ & $05.2$\\
FFHQ & $73.1$ & $17.1$ & $09.8$ &  & $64.9$ & $35.1$ &  & $\pmb{44.2}$ & $\pmb{03.9}$ & $\pmb{51.9}$ &  & $32.0$ & $68.0$ &  & $55.3$ & $31.2$ & $13.6$\\
IJB-C & $\pmb{59.3}$ & $\pmb{29.7}$ & $\pmb{10.9}$ &  & $33.9$ & $66.1$ &  & $66.9$ & $01.7$ & $31.3$ &  & $80.6$ & $19.4$ &  & $\pmb{40.5}$ & $\pmb{34.4}$ & $\pmb{25.1}$\\
CASIA-WebFace & $78.4$ & $09.8$ & $11.9$ &  & $\pmb{59.3}$ & $\pmb{40.7}$ &  & $54.9$ & $02.1$ & $42.9$ &  & $49.8$ & $50.2$ &  & $71.2$ & $17.2$ & $11.5$\\
    \midrule
    \small{Facial Benchmarks (Diversity)}\\
    \cmidrule(r){1-1}
RFW (Race-diverse) & $\pmb{46.9}$ & $\pmb{25.9}$ & $\pmb{27.2}$ &  & $\pmb{51.4}$ & $\pmb{48.6}$ &  & $74.8$ & $01.6$ & $23.6$ &  & $\pmb{48.7}$ & $\pmb{51.3}$ &  & $\pmb{69.7}$ & $\pmb{24.3}$ & $\pmb{06.0}$\\
BUPT (Race-diverse) & $52.6$ & $21.2$ & $26.1$ &  & $53.4$ & $46.6$ &  & $\pmb{67.4}$ & $\pmb{02.6}$ & $\pmb{30.1}$ &  & $46.8$ & $53.2$ &  & $82.1$ & $12.6$ & $05.3$\\
% CALFW (Age) & $75.9$ & $15.3$ & $08.8$ &  & $28.2$ & $71.8$ &  & $74.6$ & $00.7$ & $24.7$ &  & $58.2$ & $41.8$ &  & $82.0$ & $17.5$ & $00.5$\\
% AgeDB (Age) & $92.6$ & $06.7$ & $00.7$ &  & $22.9$ & $77.1$ &  & $\pmb{55.9}$ & $\pmb{01.2}$ & $\pmb{42.8}$ &  & $66.9$ & $33.1$ &  & $58.1$ & $41.5$ & $00.4$\\
% CFPFP (Pose) & $59.5$ & $26.5$ & $14.0$ &  & $57.8$ & $42.2$ &  & $67.3$ & $01.8$ & $30.9$ &  & $35.1$ & $64.9$ &  & $\pmb{48.4}$ & $\pmb{08.7}$ & $\pmb{42.9}$\\
% CPLFW (Pose) & $70.2$ & $20.8$ & $08.9$ &  & $27.4$ & $72.6$ &  & $76.9$ & $02.5$ & $20.6$ &  & $58.9$ & $41.1$ &  & $33.6$ & $26.6$ & $39.9$\\
    \midrule
    \small{Synthetic Data Sampling}\\
    \cmidrule(r){1-1}
    Rand50 & $69.8$ & $20.8$ & $09.4$ &  & $63.0$ & $37.0$ &  & $44.0$ & $04.2$ & $51.8$ &  & $26.2$ & $73.8$ &  & $\pmb{54.5}$ & $\pmb{33.0}$ & $\pmb{12.6}$\\
    QD50 (Ours) & $\pmb{56.8}$ & $\pmb{18.1}$ & $\pmb{25.2}$ &  & $\pmb{54.3}$ & $\pmb{45.7}$ &  & $\pmb{51.1}$ & $\pmb{04.3}$ & $\pmb{44.6}$ &  & $\pmb{26.6}$ & $\pmb{73.4}$ &  & $68.0$ & $26.2$ & $05.8$\\
    \bottomrule
  \end{tabular}
  \caption{Real datasets contain large imbalances across several demographic and visual attributes. 
  Proportions ($\%$) making up each attribute are listed below. The most uniform distribution of each dataset category is bolded.
  The proportions are estimated with classifiers trained on IJB-C and CelebA labels, with the computations for each attribute described in Section~\ref{results:qd}.}
  \label{tab:datasets-and-sampling}
\end{table*}

\section{Ethical Impacts and Limitations}

Facial recognition has fraught ethical impacts when misused for the purpose of mass surveillance, and we condemn this use of the technology. 
Nonetheless, facial recognition technology has been publicly adopted and is used in identity verification systems (e.g., Face ID). 
Since facial recognition is already integrated in social and institutional contexts, we argue it must be made to be performant for all users.

Facial recognition systems notoriously perform better on light-skinned individuals and worse on dark-skinned individuals~\cite{gender-shades}. 
To our knowledge, there are no publicly released facial datasets with skin tone annotations and are balanced across skin tones, which we believe to be the correct approach to assessing the fairness of those systems. 
We use the Racial Faces in-the-Wild (RFW) dataset~\cite{rfw} as a proxy for measuring and addressing this fairness: RFW is approximately balanced across skin tones, but uses the labels Caucasian and African, labeled using the Face++ API. 
We recognize the flaws in this labeling method and reinterpret these labels as \textit{light-skinned} and \textit{dark-skinned}, respectively.

Debiasing methods often use heuristics that can be useful for some contexts, but not others.
While QDGS aims to promote uniform distributions across data attributes, it may not be suited to tasks that necessitate proportionate representations (i.e., when the synthetic dataset's distribution is similar to that of the target population's).
Further, the language prompts in QDGS enhance the expressivity of desired diversity attributes, but they require careful design to avoid reinforcing linguistic biases.
Finally, for contexts that involve vectorized data or motivate human feedback, practitioners may consider using QDGS in conjunction with other methods~\cite{dpp-summarization, faircrowd, qd-feedback}.

\section{Preliminaries}

\subsection{\ifdefined\arxiv{Learning from synthetic data}\else{Learning from Synthetic Data}\fi}
Advancements in generative models enable the creation of complex and realistic synthetic data, but these improvements often depend on larger models. 
As generative models scale, their latent spaces also become increasingly complex to navigate, motivating latent space exploration algorithms as part of the sampling process to protect data diversity.

Formally, these latent space exploration algorithms identify inputs $\bm\theta\in\mathbb{R}^n$ for a generative model $\mathcal{G}$ that produce an image $X=\mathcal{G}(\bm\theta)$, optionally  maximizing an objective $f:\mathbb{R}^n\rightarrow\mathbb{R}$.
To increase sample diversity, inputs are augmented with additive sample noise: $\bm\theta^\prime=\bm\theta + z$~\cite{google-imagenet, phillip-stable}; similarly, objective functions can factor in diversity by including distance in latent space as a component of the function: $f(\bm\theta^\prime)=g(\bm\theta^\prime)+\beta||\bm\theta^\prime - \bm\theta||_2$~\cite{mcduff-bayesian, jain-pretrain}.

However, large distances in latent space may not correspond to distinct outputs, due to mode collapse in generative models~\cite{mode-collapse}.
Further, latent distances do not directly describe any specific semantic concept, so the resulting variance in data may be ineffective for the downstream task at hand.
We refer to distance in latent space as \textit{unmeasured diversity}, since the dimension of variance is ambiguous.
While unmeasured diversity is useful for promoting heterogeneity, it does not give practitioners control over the desired concepts for which data should vary.
Thus, in addition to unmeasured diversity, we argue that generative sampling algorithms should integrate methods that control \textit{measured diversity} (i.e., dimensions of diversity that are aligned with semantic concepts).
In QDGS, we estimate these dimensions with similarity scores to language prompts.

\subsection{\ifdefined\arxiv{Fairness in facial recognition}\else{Fairness in Facial Recognition}\fi}

Despite potential misuse of facial recognition technology for mass surveillance, its widespread application requires that facial recognition technology be performant for all users.
However, large-scale facial datasets tend to disproportionately represent demographic and visual features.
Labeled Faces in-the-Wild (LFW)~\cite{lfw} contains a $\sim\!8.6\!:\!1$ ratio for light to dark skin tones, a $\sim\!3.1\!:\!1$ ratio for masculine to feminine gender presentations, and a $\sim\!2.4\!:\!1$ ratio for old to young ages~(Table \ref{tab:datasets-and-sampling}).
Other facial datasets contain similar demographic imbalances, including CelebA~\cite{celeba}, FFHQ~\cite{ffhq},  IARPA Janus Benchmark-C (IJB-C)~\cite{ijb}, and CASIA-Webface (CASIA)~\cite{casia}.

To improve the transparency of biases in facial data, RFW was designed as a benchmark with separate racial evaluation splits.
Subsequently, BUPT-Balancedface (BUPT)~\cite{bupt} was released as a training dataset with racially balanced images.
While both of these datasets are estimated to have more uniform distributions of skin tones, they still have high gender presentation disparity with a masculine to feminine ratio of $\sim\!3.2\!:\!1$ in RFW and $\sim\!2.2\!:\!1$ in BUPT. 
This imbalance reflects the challenge of creating intersectional representations for training datasets.

\subsection{\ifdefined\arxiv{Quality-diversity optimization}\else{Quality-Diversity Optimization}\fi}\label{prel:qd}

Quality-diversity~\cite{qd-background} is a subfield of evolutionary optimization where $n$-dimensional solutions maximize an objective $f:\mathbb{R}^n\rightarrow\mathbb{R}$ and cover
a measure space $S$ formed by the range of $k$ measures $m_j:\mathbb{R}^n\rightarrow\mathbb{R}$, or as a joint measure, $\bm{m}:\mathbb{R}^n\rightarrow\mathbb{R}^k$.
For each $s\in S$, quality-diversity algorithms aim to find a solution $\bm{\theta}\in\mathbb{R}^n$ such that $\bm{m}(\bm{\theta})=s$ and $f(\bm\theta)$ is maximized.
In the quality-diversity literature, the task of searching over latent solutions is called latent space illumination~\cite{lsi}.

We employ CMA-MAEGA~\cite{cma-mae} to solve a relaxed version of quality-diversity by using a tessellation $T$ of measure space $S$.
First, $\bm\theta$ is initialized to an initial solution $\bm\theta_0$.
For each of $N$ iterations, a population of $\lambda$ coefficients $\bm{c}$ are sampled from a $k+1$ Gaussian distribution $\mathcal{N}(\bm\mu, \Sigma)$.
Next, solutions are branched with objective-measure gradients:
\begin{equation}
\bm\theta^\prime=\bm\theta+|\bm c_0|\bm\nabla f(\bm\theta) + \sum_{j=1}^{k}c_j\bm\nabla m_j(\bm\theta)
\end{equation}

An annealed improvement scalar $\Delta_i$ is used to rank sampled gradients $\bm\nabla_i$ for each new solution $\bm\theta^\prime$ based on the update history of a measure cell.
We perform gradient ascent on $\bm\theta$ weighted on improvement ranking $\Delta_i$.
The Gaussian parameters $\bm\mu$ and $\Sigma$ are updated with an evolutionary adaptation step.
Finally, if no improvement was found, a new seed solution $\bm\theta$ is sampled uniformly at random from existing solutions.
We visualize these steps in Fig.~\ref{fig:overview}.

\begin{algorithm}[tb]
\caption{Quality-Diversity Generative Sampling}
\label{alg:algorithm}
\textbf{Input}: A generative model $\mathcal{G}$, objective prompts $\bm{x}_f$, a regularization distance $\rho$, and measure prompts $\{\bm{x}_{m_j}\}_{j=1}^{k}$.\\
\textbf{Parameters}: 
An initial solution $\bm\theta_0$, the number of iterations $N$, a learning rate $\eta$, a branching population size $\lambda$, and an initial branching step size $\sigma_g$.\\
\textbf{Output}: A dataset $\mathcal{D}_{archive}$ with latent solutions for generative model $\mathcal{G}$.\\
\begin{algorithmic}[1] %[1] enables line numbers
%%% SOLUTION INITIALIZATION %%%
\STATE Initialize  $\mathcal{D}_{archive}=\phi$, $\bm\theta=\bm\theta_0$, $\bm\mu=\bm0$, and $\Sigma=\sigma_gI$.
\FOR{$iter\leftarrow1$ \textbf{to} $N$}
\STATE $X\leftarrow\mathcal{G}(\bm\theta)$
\STATE{
$f,\bm\nabla_f,\leftarrow\text{compute\_objective}(\bm\theta,X,\bm{x}_f,\rho)$
}
\STATE{
$\bm{m},\bm\nabla_{\bm{m}}\leftarrow\text{compute\_measures}(X,\{\bm{x}_{m_j}\}_{j=1}^k)$
}
\STATE{
    $\bm\nabla_f\leftarrow 
    \text{normalize}(\bm\nabla_f),
    \bm\nabla_{\bm{m}}\leftarrow
    \text{normalize}(\bm\nabla_{\bm{m}})$
}
\STATE{
    $\mathcal{D}_{archive}\leftarrow\text{update}(
    \mathcal{D}_{archive}, (\bm\theta, f, \bm{m}))$
}
\FOR{$i\leftarrow1$ \textbf{to} $\lambda$}
\STATE{$\bm{\theta}_i^\prime\leftarrow\text{branch}(\bm\theta,\bm\nabla_f,\bm\nabla_{\bm m},\bm\mu,\Sigma)$
}
\STATE{
    $X_i^\prime\leftarrow
    \mathcal{G}(\bm\theta_i^{\prime})$
}
\STATE{
$f^\prime,*\leftarrow\text{compute\_objective}(\bm\theta_i^\prime,X_i^\prime,\bm{x}_f,\rho)$
}
\STATE{
$\bm{m}^\prime,*\leftarrow\text{compute\_measures}(X_i^\prime,\{\bm{x}_{m_j}\}_{j=1}^k)$
}
\STATE{
    $\Delta_i\leftarrow
    \text{improvement}((f^\prime, \bm{m}^\prime), \mathcal{D}_{archive})$
}
\STATE{
    $\mathcal{D}_{archive}\leftarrow\text{update}(
    \mathcal{D}_{archive}, (\bm\theta_i^\prime, f^\prime, \bm{m}^\prime))$
}
\ENDFOR
\STATE{
$\bm\theta\leftarrow\text{ranked\_ascent}(\eta,\bm\nabla_{1\ldots\lambda},\Delta_{1\ldots\lambda})$
}
\STATE{
    $\bm\mu,\Sigma\leftarrow\text{adapt}(\Delta_{1\ldots\lambda})$
}
\IF{\textit{there is no change in} $\mathcal{D}_{archive}$}
\STATE{
    \text{Reset} $\bm\mu=0,\Sigma=\sigma_gI$
}
\STATE{
    Randomly select $\bm\theta,*\in\mathcal{D}_{archive}$
}
\ENDIF
\ENDFOR
\STATE {
    \textbf{return} $\mathcal{D}_{archive}$
}
\end{algorithmic}
\end{algorithm} 

\section{QDGS with Color-Biased Shapes}
To demonstrate QDGS (Algorithm~\ref{alg:algorithm}), we first use a proof-of-concept colored shapes dataset with a controlled gradation of bias strengths $b\in [0.80, 0.85, 0.90, 0.95, 0.98]$.
For each bias strength $b$, we generate a ``real'' training dataset such that red triangles and blue squares make up $b$ proportion of the dataset and blue triangles and red squares make up the rest.
A variational autoencoder $\mathcal{G}$ is trained on the version with $b=0.98$.
We show that despite the biased generative model $\mathcal{G}$, QDGS can produce a synthetic dataset that repairs accuracy on a balanced evaluation set with $b=0.50$.

\begin{figure}[tbp]
\centering
\begin{subfigure}[b]{0.48\columnwidth}
     \centering
     \includegraphics[width=\textwidth]{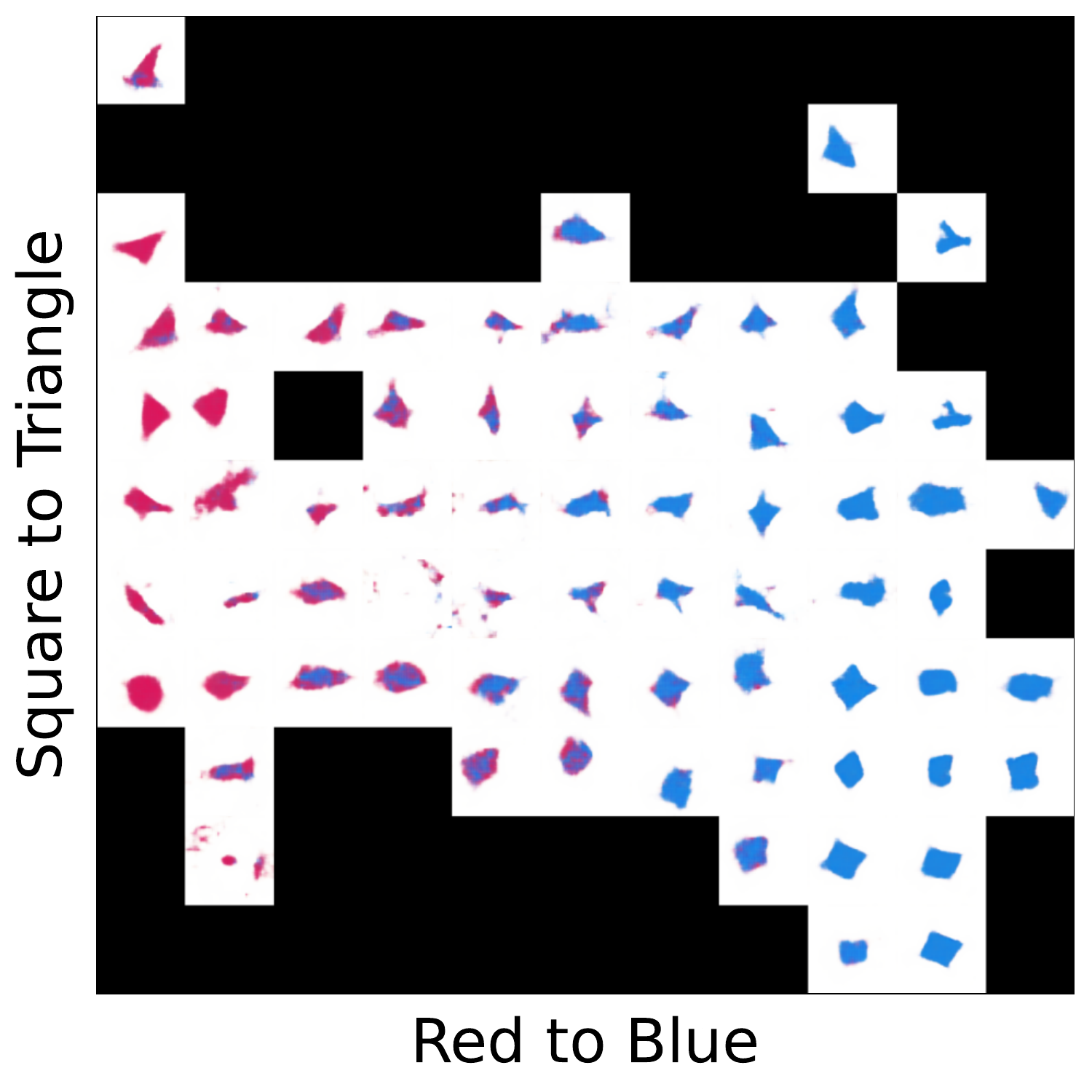}
     \caption{Random Sampling}
 \end{subfigure}
 \begin{subfigure}[b]{0.48\columnwidth}
     \centering
     \includegraphics[width=\textwidth]{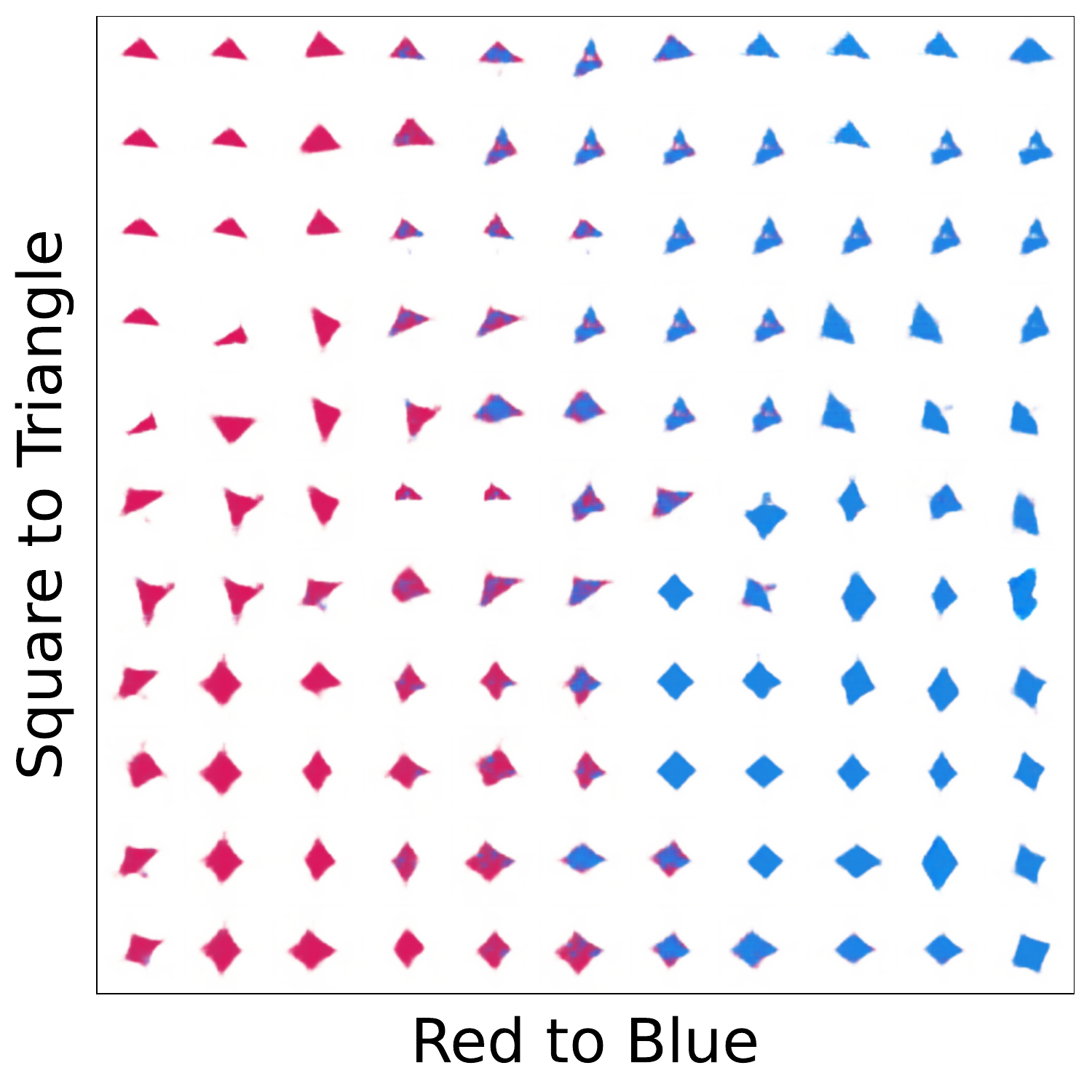}
     \caption{QDGS}
 \end{subfigure}
\caption{Despite sampling from the same biased red-triangle/blue-square generator, QDGS achieves more uniform coverage over an intersectional measure space.}
\label{fig:shape_collage}
\end{figure}

\subsection{\ifdefined\arxiv{Data generation and sampling}\else{Data Generation and Sampling}\fi}
QDGS finds solutions that maximize an objective function $f$ and measure functions $\{m_j\}_{j=1}^{k}$. 

\paragraph{Objective function:} 
The objective function $f$ computes CLIP similarity scores of generated images with positive and negative language prompts $\bm x_f=(x_f^{pos}, x_f^{neg})$:

\begin{align}
\begin{split}
f(\bm\theta)&=\cos_{\texttt{CLIP}}(\mathcal{G}(\bm\theta), x_{f}^{pos})
-\cos_{\texttt{CLIP}}(\mathcal{G}(\bm\theta), x_{f}^{neg})
\end{split}
\label{eq:clip}
\end{align}

We use the positive prompt $x_f^{pos}=$ ``\textit{A regular square or triangle}'' and to discourage out-of-distribution images, we use the negative prompt $x_f^{neg}=$ ``\textit{Splatters of colors}''.

\paragraph{Measure functions:} Each measure function $m_j$ also computes CLIP scores (Equation~\ref{eq:clip}) to determine the measures of diversity in the resulting images.
We run QDGS on $k\!=\!2$ measure dimensions, with the following prompts: $\bm{x}_{m_1}=$ ``\textit{A \{red, blue\} shape}'', $\bm{x}_{m_2}=$ ``\textit{A \{square or diamond with 4 edges, triangle with 3 edges\}}''.
We assign labels via $m_2$ with a threshold of $0.01$.

\subsection{\ifdefined\arxiv{Discriminator training and evaluation}\else{Discriminator Training and Evaluation}\fi}

For each version of our biased dataset, we train a convolutional neural network to discriminate between triangles and squares.
Optionally, we fine-tune with the dataset generated with QDGS from the biased autoencoder $\mathcal{G}$.

\begin{figure*}[tbp]
    \centering
    {
    \includegraphics[width=0.49\linewidth]{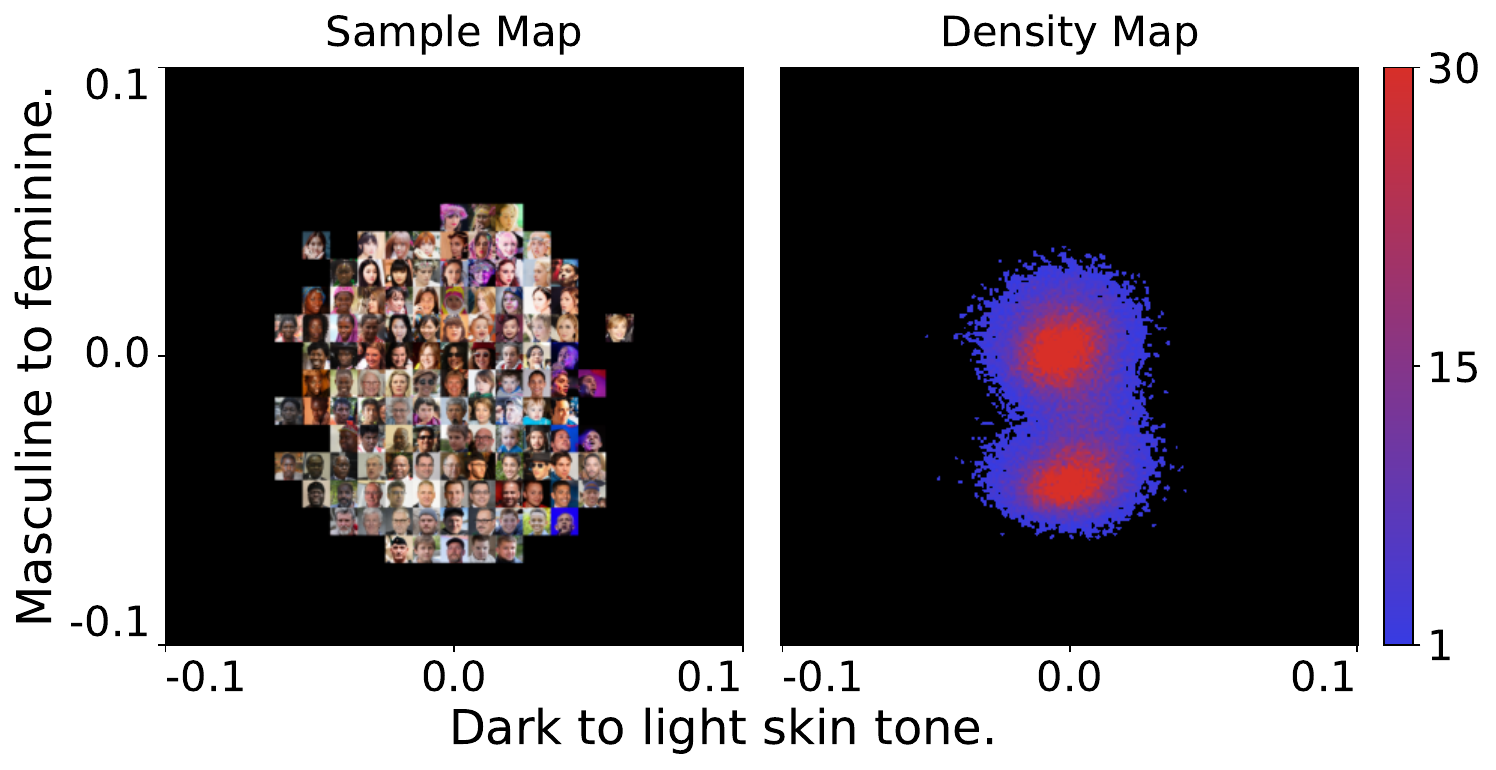}
    }
    \hfill
    {
    \includegraphics[width=0.49\linewidth]{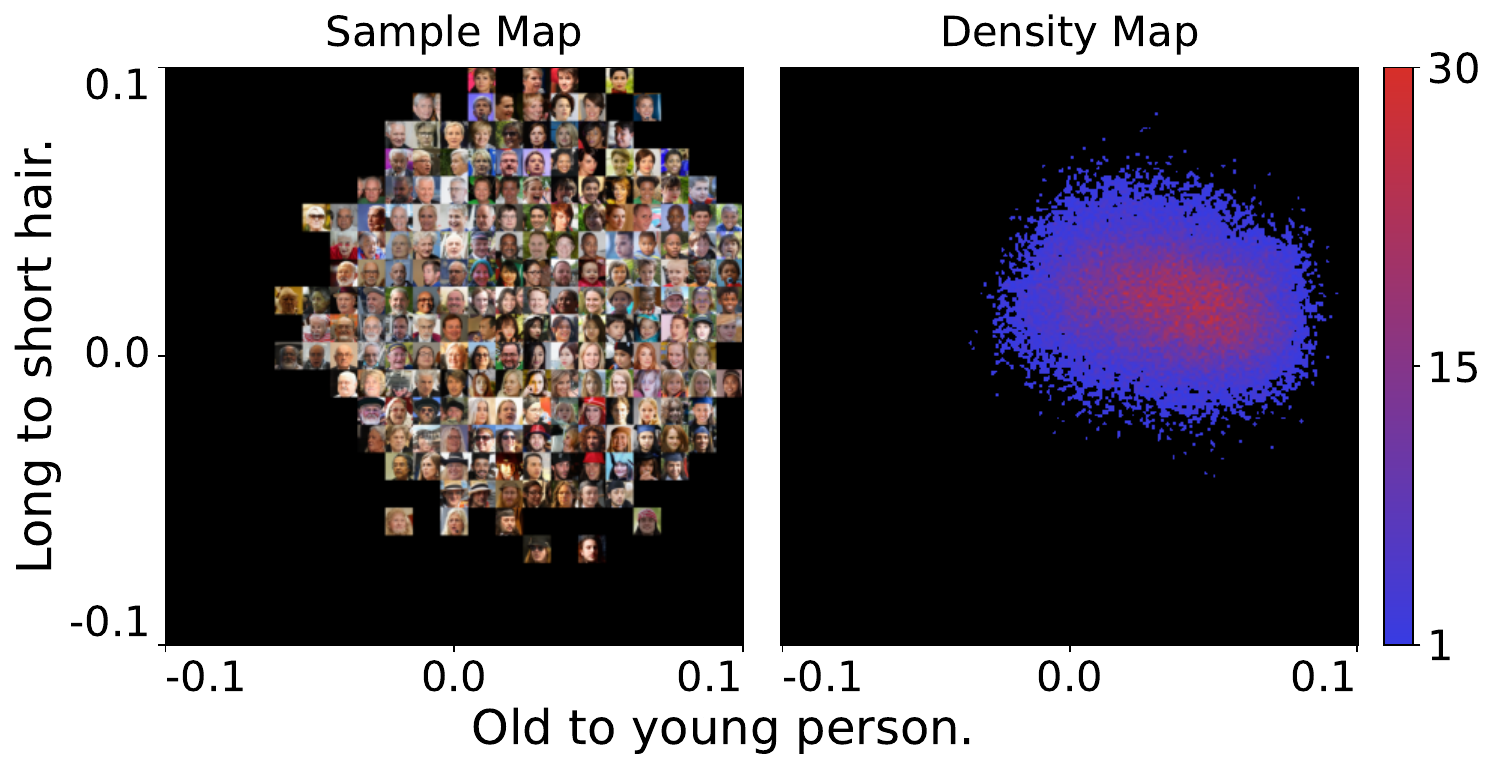}
    }\\
    {
    \includegraphics[width=0.49\linewidth]{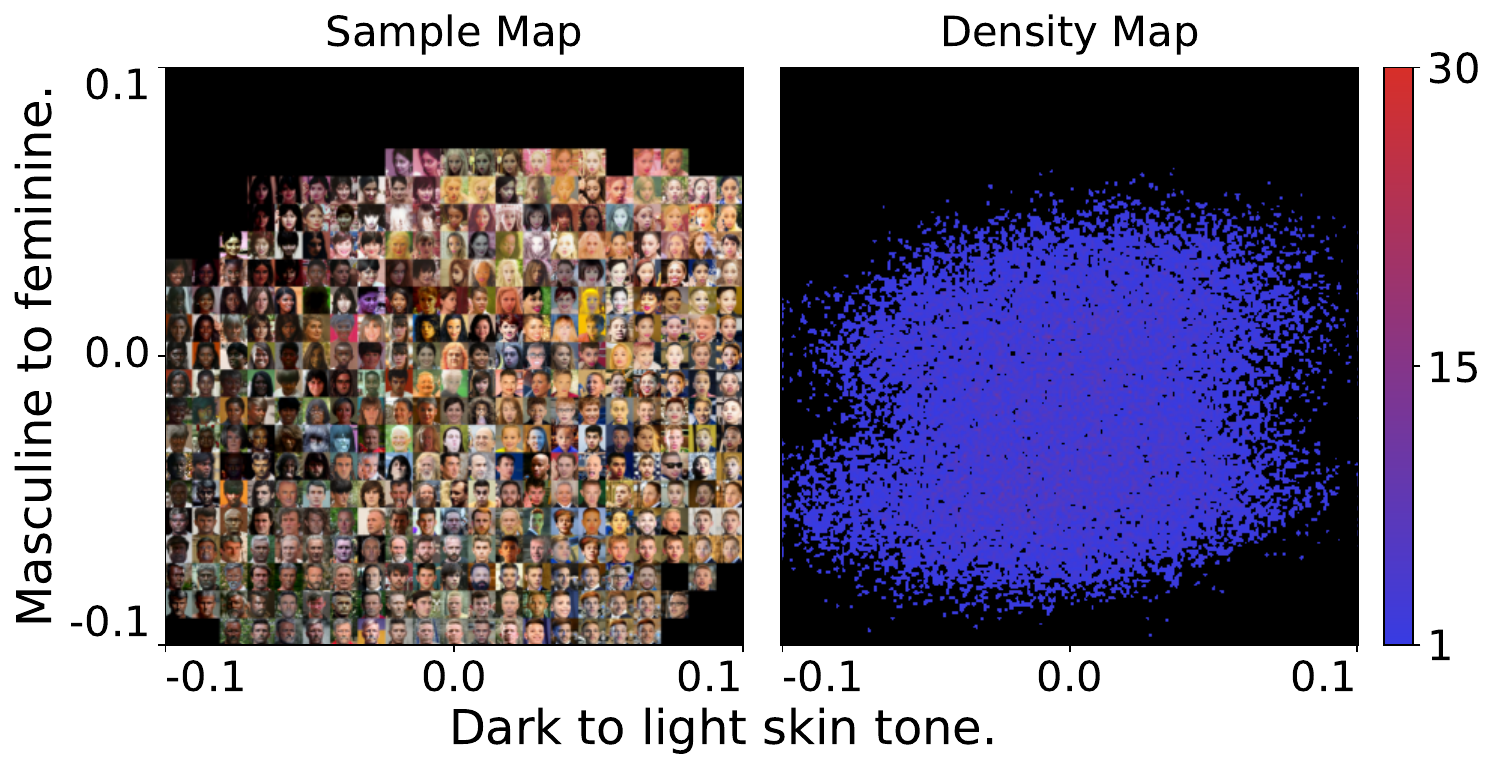}
    }
    \hfill
    {
    \includegraphics[width=0.49\linewidth]{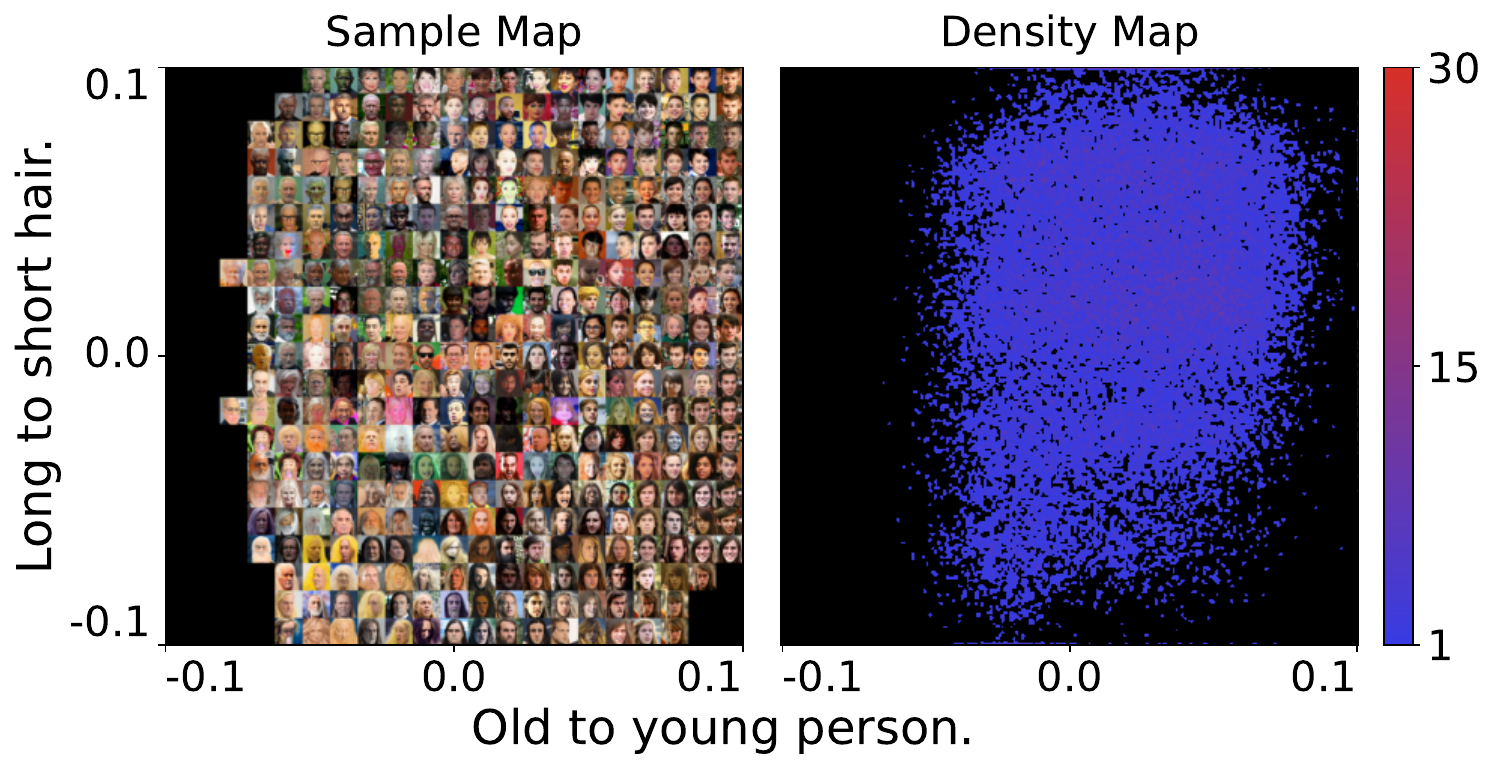}
    }
    
    \caption{
    Sample and density maps of synthetic datasets projected onto CLIP latent planes, computed via measure functions described in Section~\ref{qdgs:data-gen}.
    \ifdefined\arxiv{\textbf{Top}}\else{Top}\fi: Random sampling (Rand50) creates high-density clusters and have sparse representation of extreme measures.
    \ifdefined\arxiv{\textbf{Bottom}}\else{Bottom}\fi: QDGS (QD50) can create more uniform distributions of synthetic data across measures.}
    \label{fig:collages}
\end{figure*}

\subsection{\ifdefined\arxiv{Results for color-biased shapes}\else{Results for Color-Biased Shapes}\fi}

Applying QDGS to our proof-of-concept dataset demonstrates two key observations: (1) QDGS is able to generate uniform coverage over intersectional measure spaces; and (2) training with QDGS helps repair biases in classifiers.

In Fig.~\ref{fig:shape_collage}, we show that QDGS is able to generate a balanced proportion of colors with shapes.
When randomly sampling, the latent distribution of generator $\mathcal{G}$ spuriously associates red shapes with triangles and blue shapes with squares. 
Further, generated outputs can sometimes fall out of distribution, resulting in color splatters rather than shapes.
Conversely, sampling with QDGS is able to produce balanced images, which are often higher in quality, by evenly distributing red and blue shapes with triangles and squares.

For each bias level, classifier accuracy increases on a fair evaluation benchmark, improving accuracy by up to $\approx27\%$ (Fig.~\ref{fig:cover}).
Notably, QDGS is more helpful when biases in real datasets are stronger.
These results suggest that generative models, even when they are biased, are more helpful for training classifiers when used with QDGS.

\section{QDGS with Facial Recognition}
We generate synthetic facial images using the StyleGAN2 model $\mathcal{G}$ with a resolution of $256\times256$ trained on FFHQ.
We create two pretraining datasets with QDGS, QD15 and QD50, from 15K and 50K solutions respectively. 
We also create randomly sampled baseline datasets Rand15 and Rand50.
In all datasets, latent solutions $\bm\theta$ are sampled from StyleGAN2's larger latent space $\mathcal{W}\in\mathbb{R}^{18\times512}$.

\subsection{\ifdefined\arxiv{Data generation and sampling}\else{Data Generation and Sampling}\fi}\label{qdgs:data-gen}
\paragraph{Objective function:} 
The objective function $f$ is a weighted sum ($\beta_1=0.5,\beta_2=0.2$) of three component functions, including a text objective $g_{txt}$, an image margin $g_{mgn}$, and a regularization factor $g_{reg}$:
\begin{equation}
f(\bm\theta) =
g_{txt}(\mathcal{G}(\bm\theta))-
\beta_1 g_{mgn}(\mathcal{G}(\bm\theta))-
\beta_2 g_{reg}(\bm\theta)
\end{equation}

The text objective (Equation~\ref{eq:clip}) is used for fine-grained guidance over the semantic quality of visual features.
While synthetic image quality can be estimated with other metrics, such as Fr\'echet inception distance (FID) or a classifier trained on the \textit{Blurry} label of CelebA, these metrics typically do not describe noise structures specific to a domain-task relationship, such as occlusions of the face due to hair or a face mask.
Beyond identifying specific noise structures in a domain-task relationship, language prompts can also describe complex, high-level desiderata that would otherwise necessitate the composition of several low-level heuristics.
We use the positive prompt $x_f^{pos}=$ ``\textit{A detailed photo of an individual with diverse features}'' and to discourage out-of-distribution images, we use the negative prompt $x_f^{neg}=$ ``\textit{An obscure, fake, or discolored photo of a person}''.

We incorporate a distance margin to the objective to boost exploration of solutions via unmeasured diversity.
We maximize that margin on the subset of randomly selected solutions generated from the past $100$ iterations, which we denote $\mathcal{D}_{mem}\subset\mathcal{D}_{archive}$:
\begin{equation}
    g_{mgn}(X)=\max_{X^\prime\in \mathcal{D}_{mem}}
    \text{cos}_{\texttt{CLIP}}(X, X^\prime)
\end{equation}

Finally, we place a constraint on the domain of acceptable solutions by regularizing on the latent space $\mathcal{W}$.
We first identify the mean standardized Euclidean distance $\delta_{reg}$ from $10,000$ vectors sampled from the latent space $\mathcal{W}$.
We then penalize solutions outside distance $\delta_{reg}$, with a distance multiplier $\rho=0.5$, where solutions within the boundary distance are not penalized:
\begin{equation}
    g_{reg}(\bm\theta)=(\max
    (\delta(\bm\theta, W), \rho\delta_{reg})-\rho\delta_{reg})^2
\end{equation}

\paragraph{Measure functions:} We use Equation~\ref{eq:clip} to compute the attribute diversity in the resulting images on $k\!=\!4$ measure dimensions, with the following prompts: $\bm{x}_{m_1}=$ ``\textit{A \{feminine, masculine\} person}'', $\bm{x}_{m_2}=$ ``\textit{A person with \{light, dark\} skin}'', $\bm{x}_{m_3}=$ ``\textit{A person with \{short, long\} hair}'', and $\bm{x}_{m_4}=$ ``\textit{A person in their \{20, 50\}s}''.

\subsection{\ifdefined\arxiv{Label assignment and data augmentation}\else{Label Assignment and Data Augmentation}\fi}\label{sec:data_augmentations}

To assign labels while maintaining feasibility for memory, we first separate each dataset into equally-sized $3\!\times\!3$ chunks based on the first two measures, $m_1$ and $m_2$.
Within each chunk $\mathcal{C}_j\subset\mathcal{D}_{archive}$, we perform K-means clustering on latent solution $\bm\theta$ where $k=\frac{|\mathcal{C}_j|}{2}$, and each cluster is assigned a separate identity label.
For each latent solution $\bm\theta$, we perform latent walks along directions that correspond to horizontal facial rotation and affective valence, creating $9$ total variants of each image.

\subsection{\ifdefined\arxiv{Discriminator training and evaluation}\else{Discriminator Training and Evaluation}\fi}
We select ResNet101 backbones and train with AdaFace loss~\cite{adaface}.
For each synthetic dataset, we pretrain model weights over $26$ epochs with a learning rate of $0.01$.
Each model is trained on the skin-tone imbalanced CASIA and skin-tone balanced BUPT datasets.
We run $10$ trials for each training configuration.

To analyze facial recognition performance on skin-tone-split evaluation sets, we first evaluate on RFW.
Next, we evaluate on LFW, CFPFP~\cite{cfpfp},
CPLFW~\cite{cplfw}, CALFW~\cite{calfw}, and AgeDB~\cite{agedb}.
We follow the testing protocols of each respective benchmark.

\section{Results and Discussion}
\subsection{\ifdefined\arxiv{Spread of synthetic data distributions}\else{Spread of Synthetic Data Distributions}\fi}
QDGS is able to produce a more uniform spread of synthetic data compared to random sampling from StyleGAN2.
Randomly sampled image distributions of StyleGAN2 consist of high density clusters with one or more modes, but these modes may arise from sampling biases rather than reflecting genuine modes of real populations.
For example, while real skin tone distributions have multiple modes, randomly sampled images from StyleGAN2 tend to collapse into a single mode with respect to skin tone measures (Fig.~\ref{fig:collages}).
Conversely, QD50 exhibits higher variance and covers a broader area of the measure space compared to Rand50.
These observations are vital when training with synthetic data, as models are susceptible to overfitting on dominant classes or features.

Sampling with QDGS can also inform relationships between generative latent spaces and semantic concepts.
The marginal distributions indicate that randomly sampling $\bm\theta\in\mathcal{W}$ naturally captures a wide range of genders and ages, but underrepresented skin tones (very ``\textit{light}'' and very ``\textit{dark}'') and hair lengths (``\textit{long hair}'') are considered outliers of the distribution of $\mathcal{W}$.
Further, interactions between measures can reveal associations represented in StyleGAN2 and CLIP.
For example, image data with high similarity to language prompts (``\textit{dark skin}'' and ``\textit{feminine}''), (``\textit{light skin}'' and ``\textit{masculine}''), and (``\textit{in their 20s}'' and ``\textit{long hair}'') are sparse in QD50.
These spurious associations in synthetic datasets can amplify stereotypes in downstream tasks.

Visual inspection of the image sample maps show that images in QD50 have more extreme visual features than those in Rand50.
While these data contain more varied and possibly out-of-distribution attributes, this unmeasured diversity can help facilitate robustness and domain adaptation to unseen test cases.
For example, a model that is trained on images with neon lighting (QD50) can learn important information on lighting and colors that may help it extrapolate to predicting test cases that are in grayscale (AgeDB).

\begin{table}[tbp]
  \small
  \centering
  \setlength{\tabcolsep}{3pt}
  \begin{tabular}{l c cc c c}
    \toprule
    \textbf{Pretraining} & &
    \textbf{Dark-skinned} & \textbf{Light-skinned} & 
    \hspace{1pt} &
    \textbf{DI}
    \\
    \toprule
    \multicolumn{3}{l}{Train: CASIA (Imbalanced)}\\
    \cmidrule(r){1-3}
None &&  $88.08\pm0.07$ & $94.05\pm0.06$ &  & $93.65\pm0.10$ \\
Rand15 &&  $88.17\pm0.08$ & $94.41\pm0.07$ & & $93.38\pm0.14$ \\
Rand50 &&  $88.69\pm0.10$ & $\pmb{94.67\pm0.07}$ &  & $93.68\pm0.10$ \\
QD15 (Ours) &&  $88.25\pm0.09$ & $94.42\pm0.06$ &  & $93.47\pm0.10$ \\
QD50 (Ours) &&  $\pmb{88.94\pm0.07}$ & $94.62\pm0.10$ &  & $\pmb{93.99\pm0.06}$ \\
    \toprule
    \multicolumn{3}{l}{Train: BUPT (Balanced)}\\
    \cmidrule(r){1-3}
None &&  $96.22\pm0.03$ & $97.58\pm0.05$ &  & $\pmb{98.60\pm0.07}$ \\
Rand15 &&  $96.22\pm0.05$ & $97.72\pm0.05$ &  & $98.47\pm0.09$ \\
Rand50 &&  $96.17\pm0.09$ & $\pmb{97.82\pm0.04}$ &  & $98.31\pm0.08$ \\
QD15 (Ours) &&  $96.29\pm0.06$ & $97.75\pm0.03$ &  & $98.51\pm0.07$ \\
QD50 (Ours) &&  $\pmb{96.32\pm0.06}$ & $97.80\pm0.04$ &  & $98.49\pm0.06$ \\
    \bottomrule
  \end{tabular}
  \caption{Pretraining with QDGS can improve accuracy on the dark-skinned subgroup in RFW. Values represent the averaged accuracy ($\%$) and disparate impact ratio ($\%$) with standard error metrics, aggregated for $10$ trials.}
  \label{tab:rfw}
\end{table}
\begin{table}[tbp]
  \small
  \centering
  \setlength{\tabcolsep}{2.3pt}
  \begin{tabular}{l c ccc c c}
    \toprule
    \textbf{Tone} &
    \textbf{Baseline} &&
    \textbf{QDGS} && 
    \textbf{AdaFace} &
    \textbf{QDGS + AdaFace}
    \\
    \toprule 
    Dark &
$79.7\pm0.3$ &&  $82.8\pm0.2$ && $88.1\pm0.1$  & $\pmb{88.9}\pm\pmb{0.1}$ \\
Light &
$88.8\pm0.1$ &&  $90.8\pm0.3$ && $94.1\pm0.1$  & $\pmb{94.6}\pm\pmb{0.1}$ \\
    \bottomrule
  \end{tabular}
  \caption{Ablation results: QDGS is helpful for either loss.}
  \label{tab:rfw-ablation}
\end{table}
\begin{table*}[tbp]
  \small
  \centering
  \setlength{\tabcolsep}{5pt}
  \begin{tabular}{l c ccccccc}
    \toprule
    \textbf{Pretraining} & \hspace{20pt} &
    \textbf{LFW} & \textbf{CFPFP} & \textbf{CPLFW} & \textbf{CALFW} & \textbf{AgeDB} & \textbf{AVG}
    \\
    \toprule
    \multicolumn{2}{l}{Train: CASIA (Imbalanced)}\\
    \cmidrule(r){1-2}
None & & $99.38\pm0.02$ & $95.14\pm0.06$ & $90.24\pm0.09$ & $93.53\pm0.05$ & $94.33\pm0.06$ & $94.52\pm0.03$ \\
Rand15 & & $99.48\pm0.02$ & $95.68\pm0.05$ & $90.69\pm0.07$ & $93.62\pm0.03$ & $94.67\pm0.09$ & $94.83\pm0.03$ \\
Rand50 & & $99.45\pm0.02$ & $95.70\pm0.04$ & $90.91\pm0.08$ & $93.60\pm0.04$ & $\pmb{94.82\pm0.05}$ & $94.89\pm0.03$ \\
QD15 (Ours) & &  $99.45\pm0.02$ & $95.67\pm0.04$ & $90.78\pm0.06$ & $93.56\pm0.04$ & $94.67\pm0.06$ & $94.83\pm0.03$ \\
QD50 (Ours) & & $\pmb{99.50\pm0.01}$ & $\pmb{95.72\pm0.03}$ & $\pmb{90.94\pm0.06}$ & $\pmb{93.71\pm0.06}$ & $94.72\pm0.07$ & $\pmb{94.92\pm0.02}$ \\
    \toprule
    \multicolumn{2}{l}{Train: BUPT (Balanced)}\\
    \cmidrule(r){1-2}
None & & $99.71\pm0.02$ & $96.15\pm0.04$ & $93.00\pm0.04$ & $95.60\pm0.02$ & $96.87\pm0.05$ & $96.27\pm0.02$ \\
Rand15 & & $99.74\pm0.02$ & $\pmb{96.29\pm0.05}$ & $\pmb{93.17\pm0.05}$ & $\pmb{95.67\pm0.04}$ & $96.90\pm0.02$ & $\pmb{96.35\pm0.02}$ \\
Rand50 & & $\pmb{99.75\pm0.02}$ & $96.17\pm0.06$ & $93.07\pm0.04$ & $95.65\pm0.04$ & $\pmb{96.93\pm0.05}$ & $96.31\pm0.03$ \\
QD15 (Ours) & & $\pmb{99.75\pm0.01}$ & $96.23\pm0.04$ & $93.04\pm0.07$ & $95.66\pm0.03$ & $\pmb{96.93\pm0.05}$ & $96.32\pm0.03$ \\
QD50 (Ours) & & $99.73\pm0.02$ & $96.21\pm0.05$ & $93.08\pm0.06$ & $95.59\pm0.03$ & $96.90\pm0.07$ & $96.30\pm0.02$ \\
    \bottomrule
  \end{tabular}
  \caption{Pretraining with QDGS produces potential improvements in accuracy. Values represent averaged accuracy ($\%$) with standard error metrics evaluated on five evaluation benchmarks, aggregated for $10$ trials.}
  \label{tab:eval5sets}
\end{table*}

\subsection{\ifdefined\arxiv{Generalization of language guidance to categorical diversity labels}\else{Generalization of Language Guidance to Categorical Diversity Labels}\fi}\label{results:qd}
In addition to maximizing language-guided diversity, QDGS generalizes to unseen categorical diversity labels.
We classify data using models trained on IJB-C (\textit{Skin Tone}, \textit{Background}, \textit{Face Rotation}) and CelebA (\textit{Age}, \textit{Gender}), where each prediction probability is discretized into two (\textit{Background}, \textit{Age}) or three (\textit{Skin Tone}, \textit{Gender}, \textit{Face Rotation}) bins.
In Table~\ref{tab:datasets-and-sampling}, we present these proportions for facial benchmarks and our synthetic datasets Rand50 and QD50.

For labels corresponding to measures (e.g.,  $\textit{Skin Tone}:m_1$), QD50 is able to achieve a more uniform distribution than Rand50.
QDGS increases the proportion of images recognized as dark skin tones from $9.4\%$ to $25.2\%$.
QDGS is also able to simultaneously increase the proportion of images recognized as old from $37.0\%$ to $45.7\%$.
These results demonstrate that balancing effects from QDGS generalizes to predictive distributions for CelebA and IJB-C labels (e.g., \textit{Fitzpatrick Skin Type 6}), even though QDGS uses language prompted measures (e.g., ``\textit{dark skin}'').
Generalization to categorical diversity labels supports that language prompting can be valuable approximations of diversity without the use of additional labeled data. 

However, it is intractable to anticipate an exhaustive list of the desired diversity measures \textit{a priori}.
This challenge is demonstrated by facial rotation, which has a high skew in QD50.
QDGS also relies on a generative model's ability to generate desired concepts, which may not be the case when the data are simply not represented to a sufficient standard.
In these scenarios, QDGS will be unsuccessful at generalizing to the corresponding categorical labels.

\subsection{\ifdefined\arxiv{Repairing biases in classifiers with QDGS}\else{Repairing Biases in Classifiers with QDGS}\fi}

Using QDGS, we show that facial recognition models can be more performant on faces with dark skin tones.
We compute the accuracy on reinterpreted RFW test splits and the disparate impact ratio (DI) for dark-skinned to light-skinned data (Table~\ref{tab:rfw}).
DI is represented as the ratio of performance on a minority group to that of a majority group, where values closest to $100\%$ indicates equal treatment of groups.

Models pretrained with QD15/50 achieve higher accuracies for dark-skinned faces and the same accuracies for light-skinned faces compared to those pretrained with Rand15/50.
On models trained on skin-tone imbalanced data (CASIA), QDGS achieves the highest accuracy for dark-skinned faces, raising scores from $88.08\%$ to $88.94\%$.
On models trained on skin-tone balanced data (BUPT), pretraining with QDGS has a smaller but still noticeable improvement on dark-skinned faces, achieving the highest accuracy for dark-skinned faces across training methods.
These trends are reflected in DI, with QD15/50 consistently outperforming Rand15/50, especially when models are trained on imbalanced data.
These results indicate that QDGS is helpful for pretraining, particularly when training data are imbalanced.

AdaFace, an adaptive loss that assigns higher weight to difficult training examples, may have some debiasing properties.
To explore whether QDGS is effective without using an adaptive loss, we conduct an ablation (Table~\ref{tab:rfw-ablation}).
For our baseline, we run on the ArcFace loss~\cite{arcface}.
In both cases, using QDGS improves performance on dark-skinned faces ($79.7\%$ to $82.8\%$ on ArcFace and $88.1\%$ to $88.9\%$ on AdaFace), suggesting that QDGS is helpful when used in conjunction with other debiasing methods.

\subsection{\ifdefined\arxiv{Effects of QDGS on accuracy}\else{Effects of QDGS on Accuracy}\fi}

QDGS can maintain or improve accuracy, and we do not observe accuracy decline on any evaluation benchmark.
Since all pretrained models perform better than no pretraining, we believe that synthetic data can help to prime models with diverse data attributes.
Thus, the models are equipped to generalize across distributional discrepancies when encountering novel data, surpassing models without pretraining.

On models trained with CASIA, QDGS achieves a higher average accuracy than randomly sampled pretraining (Table~\ref{tab:eval5sets}).
These differences are larger between QD50 and Rand50 than for QD15 and Rand15.
On models trained with BUPT, there are fewer differences in accuracy based on the pretraining method.
These trends likely reflect the larger scale of BUPT ($\approx$ 1.3M images in BUPT v.s. $\approx$ 200K images in CASIA), since training on a larger dataset can diminish the effects of pretraining.
Another possibility is that BUPT contains more heterogeneous data and therefore experiences fewer added benefits from QDGS.

\section{Conclusions}
This work presents QDGS, a framework for sampling balanced synthetic training datasets from generative models.
QDGS can increase spread over desired attributes, described by user-defined language prompts, by exploring high dimensionality latent spaces.
With QDGS, we are able to repair biases in shape classifiers up to $\approx27\%$.
In the more challenging domain of facial recognition, we are able to achieve a more uniform spread of synthetic facial image training data, simultaneously increasing the proportion of images recognized with dark skin tones from $9.4\%$ to $25.2\%$ and images recognized as old from $36.0\%$ to $45.7\%$.
For models trained on biased datasets, QDGS improves accuracy performance on dark-skinned faces from $88.08\%$ to $88.94\%$, and QDGS achieves the highest average accuracy across facial recognition benchmarks.
These results support training with balanced synthetic datasets. 

\section*{Acknowledgements}
This product contains or makes use of the following data made available by the Intelligence Advanced Research Projects Activity (IARPA): IARPA Janus Benchmark C (IJB-C) data detailed at Face Challenges homepage.
The authors acknowledge the Center for Advanced Research Computing (CARC) at the University of Southern California for providing computing resources that have contributed to the research results reported within this publication. URL: \ifdefined\arxiv{\href{https://carc.usc.edu}{\texttt{https://carc.usc.edu}}}\else{https://carc.usc.edu}\fi.

\pagebreak
\bibliography{Refs/methods, Refs/generative-models, Refs/synthetic-learning, Refs/benchmarks, Refs/quality-diversity}

\begin{thebibliography}{31}
\providecommand{\natexlab}[1]{#1}

\bibitem[{Azizi et~al.(2023)Azizi, Kornblith, Saharia, Norouzi, and
  Fleet}]{google-imagenet}
Azizi, S.; Kornblith, S.; Saharia, C.; Norouzi, M.; and Fleet, D.~J. 2023.
\newblock Synthetic data from diffusion models improves imagenet
  classification.
\newblock \emph{arXiv preprint arXiv:2304.08466}.

\bibitem[{Balestriero, Bottou, and LeCun(2022)}]{lecun-bias}
Balestriero, R.; Bottou, L.; and LeCun, Y. 2022.
\newblock The effects of regularization and data augmentation are class
  dependent.
\newblock \emph{Neural Information Processing Systems (NeurIPS)}.

\bibitem[{Bau et~al.(2019)Bau, Zhu, Wulff, Peebles, Strobelt, Zhou, and
  Torralba}]{mode-collapse}
Bau, D.; Zhu, J.-Y.; Wulff, J.; Peebles, W.; Strobelt, H.; Zhou, B.; and
  Torralba, A. 2019.
\newblock Seeing what a gan cannot generate.
\newblock In \emph{International Conference on Computer Vision (ICCV)}.

\bibitem[{Buolamwini and Gebru(2018)}]{gender-shades}
Buolamwini, J.; and Gebru, T. 2018.
\newblock Gender shades: Intersectional accuracy disparities in commercial
  gender classification.
\newblock In \emph{Fairness, Accountability and Transparency (FAccT)}.

\bibitem[{Celis et~al.(2018)Celis, Keswani, Straszak, Deshpande, Kathuria, and
  Vishnoi}]{dpp-summarization}
Celis, E.; Keswani, V.; Straszak, D.; Deshpande, A.; Kathuria, T.; and Vishnoi,
  N. 2018.
\newblock Fair and diverse DPP-based data summarization.
\newblock In \emph{International Conference on Machine Learning (ICML)}.

\bibitem[{Chatzilygeroudis et~al.(2021)Chatzilygeroudis, Cully, Vassiliades,
  and Mouret}]{qd-background}
Chatzilygeroudis, K.; Cully, A.; Vassiliades, V.; and Mouret, J.-B. 2021.
\newblock Quality-Diversity Optimization: a novel branch of stochastic
  optimization.
\newblock In \emph{Black Box Optimization, Machine Learning, and No-Free Lunch
  Theorems}. Springer.

\bibitem[{Cho, Zala, and Bansal(2023)}]{dalle-bias}
Cho, J.; Zala, A.; and Bansal, M. 2023.
\newblock Dall-eval: Probing the reasoning skills and social biases of
  text-to-image generation models.
\newblock In \emph{International Conference on Computer Vision (ICCV)}.

\bibitem[{Deng et~al.(2019)Deng, Guo, Xue, and Zafeiriou}]{arcface}
Deng, J.; Guo, J.; Xue, N.; and Zafeiriou, S. 2019.
\newblock Arcface: Additive angular margin loss for deep face recognition.
\newblock In \emph{Computer Vision and Pattern Recognition (CVPR)}.

\bibitem[{Ding et~al.(2023)Ding, Zhang, Clune, Spector, and
  Lehman}]{qd-feedback}
Ding, L.; Zhang, J.; Clune, J.; Spector, L.; and Lehman, J. 2023.
\newblock Quality Diversity through Human Feedback.
\newblock \emph{arXiv preprint arXiv:2310.12103}.

\bibitem[{Fontaine and Nikolaidis(2023)}]{cma-mae}
Fontaine, M.; and Nikolaidis, S. 2023.
\newblock Covariance matrix adaptation map-annealing.
\newblock In \emph{Genetic and Evolutionary Computation Conference (GECCO)}.

\bibitem[{Fontaine et~al.(2021)Fontaine, Liu, Khalifa, Modi, Togelius, Hoover,
  and Nikolaidis}]{lsi}
Fontaine, M.~C.; Liu, R.; Khalifa, A.; Modi, J.; Togelius, J.; Hoover, A.~K.;
  and Nikolaidis, S. 2021.
\newblock Illuminating mario scenes in the latent space of a generative
  adversarial network.
\newblock In \emph{AAAI Conference on Artificial Intelligence}.

\bibitem[{Friedrich et~al.(2023)Friedrich, Schramowski, Brack, Struppek,
  Hintersdorf, Luccioni, and Kersting}]{fair-diffusion}
Friedrich, F.; Schramowski, P.; Brack, M.; Struppek, L.; Hintersdorf, D.;
  Luccioni, S.; and Kersting, K. 2023.
\newblock Fair diffusion: Instructing text-to-image generation models on
  fairness.
\newblock \emph{arXiv preprint arXiv:2302.10893}.

\bibitem[{Huang et~al.(2007)Huang, Ramesh, Berg, and Learned-Miller}]{lfw}
Huang, G.~B.; Ramesh, M.; Berg, T.; and Learned-Miller, E. 2007.
\newblock Labeled Faces in the Wild: A Database for Studying Face Recognition
  in Unconstrained Environments.
\newblock Technical report, University of Massachusetts, Amherst.

\bibitem[{Jain, Memon, and Togelius(2023)}]{jain-pretrain}
Jain, A.; Memon, N.; and Togelius, J. 2023.
\newblock Zero-shot racially balanced dataset generation using an existing
  biased StyleGAN2.
\newblock \emph{arXiv preprint arXiv:2305.07710}.

\bibitem[{Jain et~al.(2022)Jain, Olmo, Sengupta, Manikonda, and
  Kambhampati}]{bias-exacerbate}
Jain, N.; Olmo, A.; Sengupta, S.; Manikonda, L.; and Kambhampati, S. 2022.
\newblock Imperfect ImaGANation: Implications of GANs exacerbating biases on
  facial data augmentation and snapchat face lenses.
\newblock \emph{Artificial Intelligence}.

\bibitem[{Karras, Laine, and Aila(2019)}]{ffhq}
Karras, T.; Laine, S.; and Aila, T. 2019.
\newblock A style-based generator architecture for generative adversarial
  networks.
\newblock In \emph{Computer Vision and Pattern Recognition (CVPR)}.

\bibitem[{Karras et~al.(2020)Karras, Laine, Aittala, Hellsten, Lehtinen, and
  Aila}]{stylegan}
Karras, T.; Laine, S.; Aittala, M.; Hellsten, J.; Lehtinen, J.; and Aila, T.
  2020.
\newblock Analyzing and improving the image quality of stylegan.
\newblock In \emph{Computer Vision and Pattern Recognition (CVPR)}.

\bibitem[{Kim, Jain, and Liu(2022)}]{adaface}
Kim, M.; Jain, A.~K.; and Liu, X. 2022.
\newblock Adaface: Quality adaptive margin for face recognition.
\newblock In \emph{Computer Vision and Pattern Recognition (CVPR)}.

\bibitem[{Kou et~al.(2021)Kou, Zhang, Shang, and Wang}]{faircrowd}
Kou, Z.; Zhang, Y.; Shang, L.; and Wang, D. 2021.
\newblock Faircrowd: Fair human face dataset sampling via batch-level
  crowdsourcing bias inference.
\newblock In \emph{International Workshop on Quality of Service (IWQOS)}.

\bibitem[{Liu et~al.(2015)Liu, Luo, Wang, and Tang}]{celeba}
Liu, Z.; Luo, P.; Wang, X.; and Tang, X. 2015.
\newblock Deep Learning Face Attributes in the Wild.
\newblock In \emph{International Conference on Computer Vision (ICCV)}.

\bibitem[{Maze et~al.(2018)Maze, Adams, Duncan, Kalka, Miller, Otto, Jain,
  Niggel, Anderson, Cheney et~al.}]{ijb}
Maze, B.; Adams, J.; Duncan, J.~A.; Kalka, N.; Miller, T.; Otto, C.; Jain,
  A.~K.; Niggel, W.~T.; Anderson, J.; Cheney, J.; et~al. 2018.
\newblock Iarpa janus benchmark-c: Face dataset and protocol.
\newblock In \emph{International Conference on Biometrics (ICB)}.

\bibitem[{McDuff et~al.(2019)McDuff, Ma, Song, and Kapoor}]{mcduff-bayesian}
McDuff, D.; Ma, S.; Song, Y.; and Kapoor, A. 2019.
\newblock Characterizing bias in classifiers using generative models.
\newblock \emph{Neural Information Processing Systems (NeurIPS)}.

\bibitem[{Moschoglou et~al.(2017)Moschoglou, Papaioannou, Sagonas, Deng,
  Kotsia, and Zafeiriou}]{agedb}
Moschoglou, S.; Papaioannou, A.; Sagonas, C.; Deng, J.; Kotsia, I.; and
  Zafeiriou, S. 2017.
\newblock Agedb: the first manually collected, in-the-wild age database.
\newblock In \emph{Faces ``In-The-Wild'' Workshop-Challenge, Computer Vision
  and Pattern Recognition (CVPR)}.

\bibitem[{Otterbacher(2018)}]{social-bias}
Otterbacher, J. 2018.
\newblock Social cues, social biases: stereotypes in annotations on people
  images.
\newblock In \emph{Human Computation and Crowdsourcing (HCOMP)}.

\bibitem[{Sengupta et~al.(2016)Sengupta, Chen, Castillo, Patel, Chellappa, and
  Jacobs}]{cfpfp}
Sengupta, S.; Chen, J.-C.; Castillo, C.; Patel, V.~M.; Chellappa, R.; and
  Jacobs, D.~W. 2016.
\newblock Frontal to profile face verification in the wild.
\newblock In \emph{Winter Conference on Applications of Computer Vision
  (WACV)}.

\bibitem[{Tian et~al.(2023)Tian, Fan, Isola, Chang, and
  Krishnan}]{phillip-stable}
Tian, Y.; Fan, L.; Isola, P.; Chang, H.; and Krishnan, D. 2023.
\newblock StableRep: Synthetic Images from Text-to-Image Models Make Strong
  Visual Representation Learners.
\newblock \emph{arXiv preprint arXiv:2306.00984}.

\bibitem[{Wang et~al.(2019)Wang, Deng, Hu, Tao, and Huang}]{rfw}
Wang, M.; Deng, W.; Hu, J.; Tao, X.; and Huang, Y. 2019.
\newblock Racial Faces in the Wild: Reducing Racial Bias by Information
  Maximization Adaptation Network.
\newblock In \emph{International Conference on Computer Vision (ICCV)}.

\bibitem[{Wang, Zhang, and Deng(2022)}]{bupt}
Wang, M.; Zhang, Y.; and Deng, W. 2022.
\newblock Meta Balanced Network for Fair Face Recognition.
\newblock \emph{Transactions on Pattern Analysis and Machine Intelligence}.

\bibitem[{Yi et~al.(2014)Yi, Lei, Liao, and Li}]{casia}
Yi, D.; Lei, Z.; Liao, S.; and Li, S.~Z. 2014.
\newblock Learning face representation from scratch.
\newblock \emph{arXiv preprint arXiv:1411.7923}.

\bibitem[{Zheng and Deng(2018)}]{cplfw}
Zheng, T.; and Deng, W. 2018.
\newblock Cross-pose LFW: A database for studying cross-pose face recognition
  in unconstrained environments.
\newblock Technical Report 18-01, Beijing University of Posts and
  Telecommunications.

\bibitem[{Zheng, Deng, and Hu(2017)}]{calfw}
Zheng, T.; Deng, W.; and Hu, J. 2017.
\newblock Cross-age lfw: A database for studying cross-age face recognition in
  unconstrained environments.
\newblock \emph{arXiv preprint arXiv:1708.08197}.

\end{thebibliography}

\ifdefined\arxiv{\appendix
\newpage

\section{Color-Biased Shapes Implementation}

\subsection{Domain details}

The real distribution of color-biased shapes consists of images $Y\in\mathbb{R}^{128\times128\times3}$.
Each image contains a red triangle or blue square with $b$ probability, or a blue triangle or red square with $1-b$ probability.
Each shape is randomly rotated from $0$ to $359$ degrees and randomly scaled from $50\%$ to $200\%$.
Visualizations of each shape are shown in Fig.~\ref{fig:appendix_shapes}:

\begin{figure}[!ht]
\centering
\includegraphics[width=0.9\columnwidth]{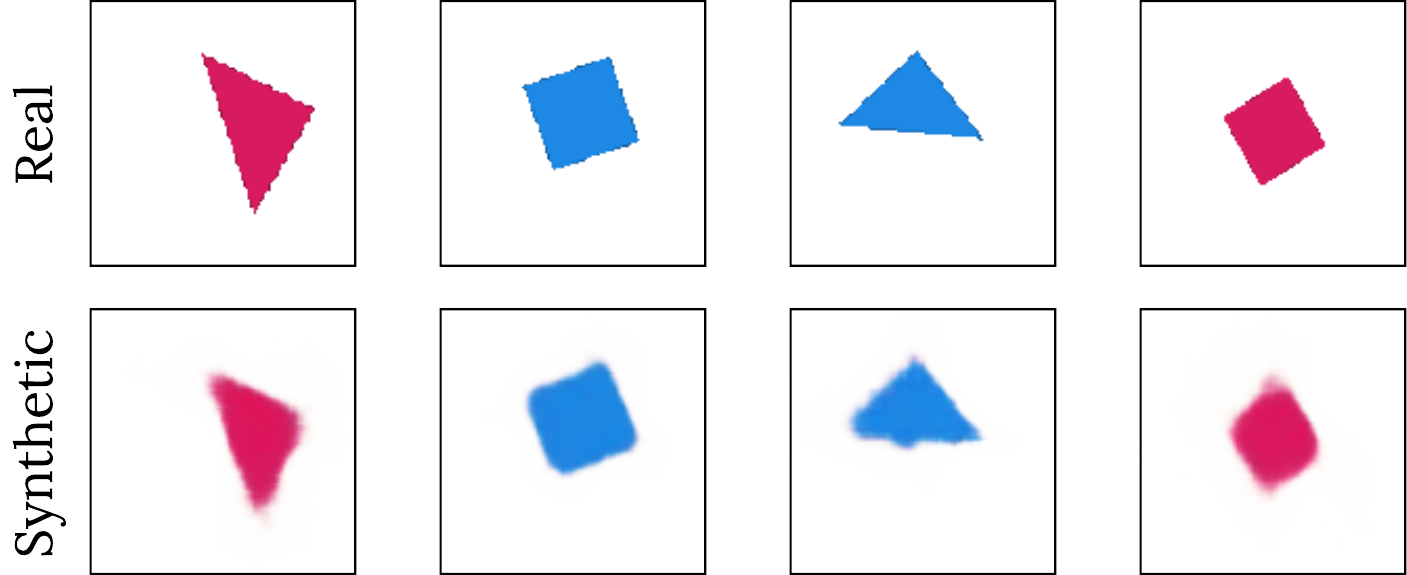}
\caption{Domain visualization of color-biased shapes. \textbf{Top}: Images randomly sampled from the real distribution. \textbf{Bottom}: Reconstructions of each image with the autoencoder.}
\label{fig:appendix_shapes}
\end{figure}

\subsection{Model architectures}

\paragraph{Shape generator:}
The generator $\mathcal{G}$ is a variational autoencoder with a $6$-dimensional embedding space. 
Reconstructed images with the generator are visualized in Fig.~\ref{fig:appendix_shapes}.

The encoder architecture comprises $3$ convolutional layers: a $3$-to-$16$, a $16$-to-$32$, and a $32$-to-$32$.
Each layer employs a ReLU activation.
The features are flattened and fed through two linear layers: the first computes the mean $\mu$ of the latent space, and the second computes the logarithm of the variance $\log(\sigma^2)$.

The decoder architecture reverses the encoding and comprises of a $8192$-dimensional linear layer followed by $3$ deconvolutional layers: a $32$-to-$32$, a $32$-to-$16$, and a $16$-to-$3$.
Each layer employs a ReLU activation, and the final output is sigmoid-activated to produce pixel values.

\paragraph{Shape classifier:}
The classifier architecture comprises $3$ convolutional layers: a $3$-to-$16$, a $16$-to-$32$, and a $32$-to-$64$. 
Each layer employs a ReLU activation, followed by $2\times 2$ max-pooling for downsampling. 
The features are flattened and fed through two linear layers: the first is a $128$-dimensional with a ReLU activation, and the second is a final sigmoid-activated linear for classification.

\section{Hyperparameter Selection}

\subsection{Objective language prompts}
Objective language prompts aim to increase the quality of the generated image, and they can be used to pre-emptively avoid failure modes in generative models.
However, seemingly harmless objective prompts that aim to enhance visual quality can result in images with features beyond their prototypical interpretations, likely due to linguistic biases.
We discuss both of these properties with respect to the shapes and facial recognition domains respectively.

\paragraph{Avoiding failure modes in generative models:}
Generative models sometimes produce out-of-distribution images.
We find that the shape generator sometimes produces blank images, and therefore we refine our objective negative prompt to be $x_f^{neg}=$``\textit{An empty image}.''
However, further inspection of the solution archive reveals that a more common failure mode of the generative model exists, particularly with images containing incoherent shapes resembling paint splatters.
Thus, we adjust the negative objective prompt to be $x_f^{neg}=$``\textit{Splatters of colors}.''

We highlight some common failure modes of the shape generator and potential positive and negative objective prompt pairings to account for them below:

\begin{figure}[!ht]
\centering
\includegraphics[width=0.9\columnwidth]{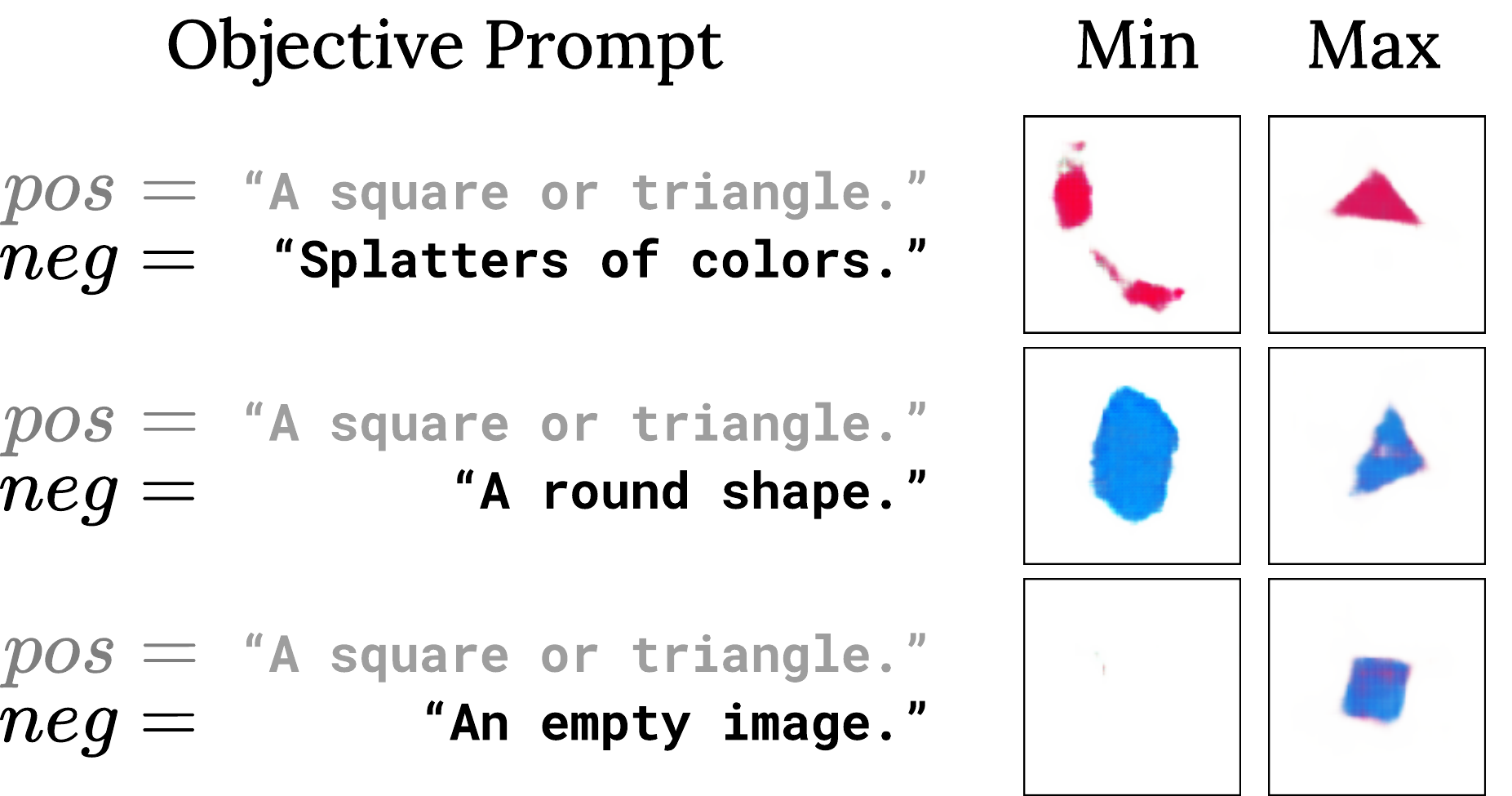}
\caption{We explicitly define objective formulations to avoid identified failure modes. For a given solution archive, the solution with the minimum and maximum objective with the corresponding prompts are shown.}
\label{fig:appendix_objectives}
\end{figure}

\paragraph{Adjusting for semantic biases:}
In our experiments, we try a variety of different adjectives to enhance the quality of StyleGAN2-generated images.
However, for certain adjectives, we find that visual features beyond the adjectives' immediate meanings are amplified:

\begin{itemize}
\item ``\textit{Well-lit}'':
produces faces with lighter skin tones in addition to higher image brightness.
\item ``\textit{HD}'': 
produces image backgrounds with photographic bokeh lighting effects.
\item ``\textit{Photorealistic}'':
produces faces that appear illustrated, with \textit{fewer} photographic qualities.
\end{itemize}

These associations demonstrate the capacity of language prompts to introduce biases or alterations that are not immediately apparent. 
In the case of generating sensitive data like facial images, these biases can have far-reaching implications.
To address these dangers, we recommend the proactive and explicit inclusion of diversity-affirmative terminology in language prompts: an example is the prompt, ``\textit{A detailed photo of an individual with \underline{diverse features}}.''
This language is intentionally ambiguous to capture dimensions of diversity that may not be considered \textit{a priori}.

\begin{table*}[tbp]
  \centering
  \setlength{\tabcolsep}{2pt}
  \begin{tabular}{l ccc c cc c ccc c cc c ccc}
    \toprule
    \textbf{Dataset} &
    \multicolumn{3}{c}{\textbf{Skin Tone}} &&
    \multicolumn{2}{c}{\textbf{Age}} &&
    \multicolumn{3}{c}{\textbf{Gender}} &&
    \multicolumn{2}{c}{\textbf{Background}} &&
    \multicolumn{3}{c}{\textbf{Face Rotation}}
    \\
    &
    \textit{\small{Light}} & \textit{\small{Mixed}} & \textit{\small{Dark}} & \hspace{2pt} & 
    \textit{\small{Young}} & \textit{\small{Old}} & 
    \hspace{2pt} &
    \textit{\small{Masc}} & \textit{\small{Andro}} & \textit{\small{Fem}} & \hspace{2pt} & 
    \textit{\small{Indoors}} & \textit{\small{Outdoors}} & 
    \hspace{2pt} &
    \textit{\small{Frontal}} & \textit{\small{Turned}} & 
    \textit{\small{Profile}}
    \\
    \toprule
    \small{Facial Benchmarks (Large Scale)}\\
    \cmidrule(r){1-1}
LFW & $72.7$ & $18.8$ & $08.5$ &  & $29.6$ & $70.4$ &  & $75.0$ & $01.0$ & $24.0$ &  & $60.2$ & $39.9$ &  & $74.2$ & $23.9$ & $01.9$\\
CelebA & $71.5$ & $20.0$ & $08.5$ &  & $70.9$ & $29.1$ &  & $41.4$ & $00.5$ & $58.2$ &  & $\pmb{49.9}$ & $\pmb{50.1}$ &  & $53.6$ & $41.2$ & $05.2$\\
FFHQ & $73.1$ & $17.1$ & $09.8$ &  & $64.9$ & $35.1$ &  & $\pmb{44.2}$ & $\pmb{03.9}$ & $\pmb{51.9}$ &  & $32.0$ & $68.0$ &  & $55.3$ & $31.2$ & $13.6$\\
IJB-C & $\pmb{59.3}$ & $\pmb{29.7}$ & $\pmb{10.9}$ &  & $33.9$ & $66.1$ &  & $66.9$ & $01.7$ & $31.3$ &  & $80.6$ & $19.4$ &  & $\pmb{40.5}$ & $\pmb{34.4}$ & $\pmb{25.1}$\\
CASIA-WebFace & $78.4$ & $09.8$ & $11.9$ &  & $\pmb{59.3}$ & $\pmb{40.7}$ &  & $54.9$ & $02.1$ & $42.9$ &  & $49.8$ & $50.2$ &  & $71.2$ & $17.2$ & $11.5$\\
    \midrule
    \small{Facial Benchmarks (Diversity)}\\
    \cmidrule(r){1-1}
RFW (Race-diverse) & $\pmb{46.9}$ & $\pmb{25.9}$ & $\pmb{27.2}$ &  & $\pmb{51.4}$ & $\pmb{48.6}$ &  & $74.8$ & $01.6$ & $23.6$ &  & $\pmb{48.7}$ & $\pmb{51.3}$ &  & $69.7$ & $24.3$ & $06.0$\\
BUPT (Race-diverse) & $52.6$ & $21.2$ & $26.1$ &  & $53.4$ & $46.6$ &  & $67.4$ & $02.6$ & $30.1$ &  & $46.8$ & $53.2$ &  & $82.1$ & $12.6$ & $05.3$\\
\rowcolor{gray!30} CALFW (Age-diverse) & $75.9$ & $15.3$ & $08.8$ &  & $28.2$ & $71.8$ &  & $74.6$ & $00.7$ & $24.7$ &  & $58.2$ & $41.8$ &  & $82.0$ & $17.5$ & $00.5$\\
\rowcolor{gray!30} AgeDB (Age-diverse) & $92.6$ & $06.7$ & $00.7$ &  & $22.9$ & $77.1$ &  & $\pmb{55.9}$ & $\pmb{01.2}$ & $\pmb{42.8}$ &  & $66.9$ & $33.1$ &  & $58.1$ & $41.5$ & $00.4$\\
\rowcolor{gray!30} CFPFP (Pose-diverse) & $59.5$ & $26.5$ & $14.0$ &  & $57.8$ & $42.2$ &  & $67.3$ & $01.8$ & $30.9$ &  & $35.1$ & $64.9$ &  & $\pmb{48.4}$ & $\pmb{08.7}$ & $\pmb{42.9}$\\
\rowcolor{gray!30} CPLFW (Pose-diverse) & $70.2$ & $20.8$ & $08.9$ &  & $27.4$ & $72.6$ &  & $76.9$ & $02.5$ & $20.6$ &  & $58.9$ & $41.1$ &  & $33.6$ & $26.6$ & $39.9$\\
    \midrule
    \small{Synthetic Data Sampling}\\
    \cmidrule(r){1-1}
    \rowcolor{gray!30} Rand15 & $69.4$ & $21.1$ & $09.5$ &  & $63.0$ & $37.0$ &  & $44.7$ & $04.1$ & $51.2$ &  & $26.3$ & $73.7$ &  & $\pmb{55.0}$ & $\pmb{33.4}$ & $\pmb{11.6}$\\
    Rand50 & $69.8$ & $20.8$ & $09.4$ &  & $63.0$ & $37.0$ &  & $44.0$ & $04.2$ & $51.8$ &  & $26.2$ & $73.8$ &  & $54.5$ & $33.0$ & $12.6$\\
    \rowcolor{gray!30} QD15 (Ours) & $\pmb{56.7}$ & $\pmb{18.1}$ & $\pmb{25.2}$ &  & $\pmb{53.8}$ & $\pmb{46.2}$ &  & $51.3$ & $04.2$ & $44.5$ &  & $26.8$ & $73.2$ &  & $67.5$ & $26.7$ & $05.8$\\
    QD50 (Ours) & $56.8$ & $18.1$ & $25.2$ &  & $54.3$ & $45.7$ &  & $\pmb{51.1}$ & $\pmb{04.3}$ & $\pmb{44.6}$ &  & $\pmb{26.6}$ & $\pmb{73.4}$ &  & $68.0$ & $26.2$ & $05.8$\\
    \bottomrule
  \end{tabular}
  \caption{Demographic and visual attribute analysis on benchmarks, in addition to those reported in Table~\ref{tab:datasets-and-sampling}. Datasets that are not previously reported are highlighted in gray. Most other benchmarks exhibit the same imbalances as other datasets, and proportions of each attribute are generally consistent for each sampling method.}
  \label{tab:datasets-and-sampling-full}
\end{table*}

\subsection{Measure language prompts}
We often find that detailed or descriptive phrasing of language prompts produce smoother measure spaces, which improves the balancing of synthetic data.
This finding is consistent with \cite{cma-mae}.

\paragraph{Shapes domain:} 
We try two variants of the measures' language prompt phrasing with varied detail:
\begin{itemize}
\item 
$\bm x_{m_1}=$``\textit{\{red, blue\}}.''
$\bm x_{m_2}=$``\textit{A \{square, triangle\}}.''
\item 
$\bm x_{m_1}=$``\textit{A \{red, blue\} shape}.''
$\bm x_{m_2}=$``\textit{A \{square or diamond with 4 edges, triangle with 3 edges\}}.''
\end{itemize}

We find that the less detailed measure prompts result in non-smooth measure values and can have more trouble covering a measure space.
For a given archive, we visualize the effects of the measure functions below:

\begin{figure}[!ht]
\centering
{
\includegraphics[width=0.48\columnwidth]{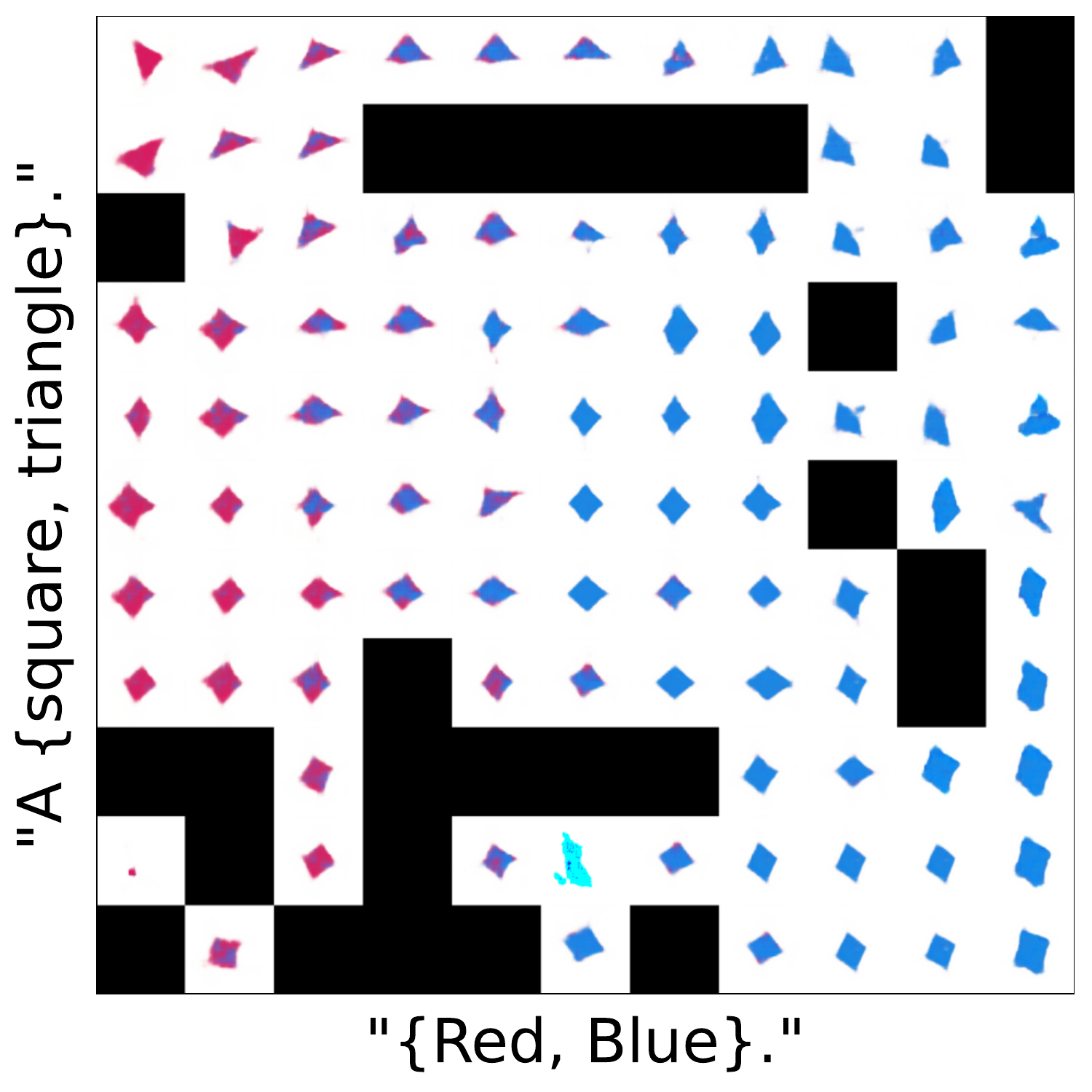}
\hfill
\includegraphics[width=0.48\columnwidth]{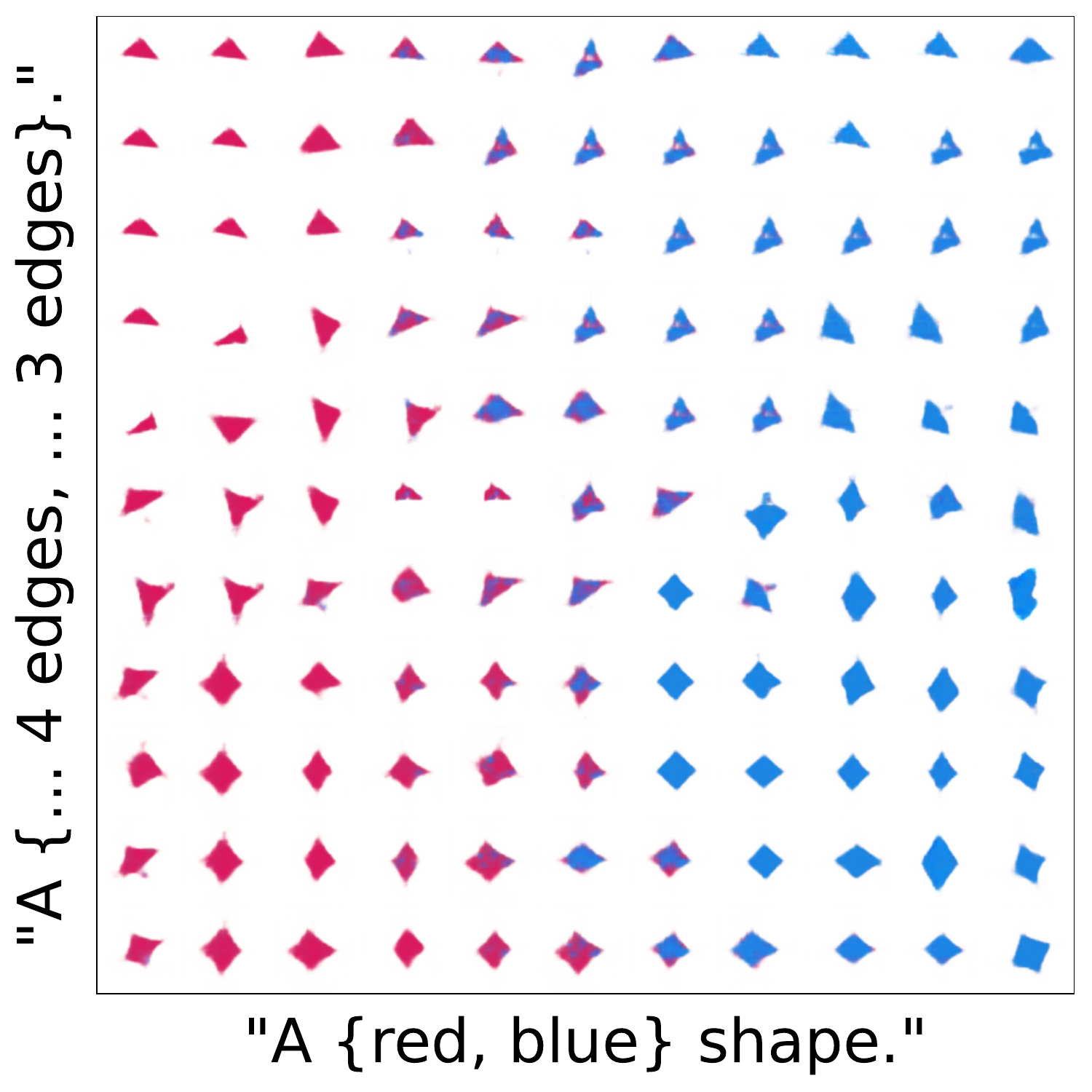}
}
\caption{Detailed measure prompts produce smooth measure values. \textbf{Left}: The archive is mapped with less detailed measure prompts. \textbf{Right}: The archive is mapped with detailed measure prompts.}
\label{fig:appendix_measure_functions}
\end{figure}

\paragraph{Facial recognition domain:} 
We choose measures that are believed to be visually salient and are also related to disparities that may affect facial recognition performance.

With respect to gender presentation, skin tone, and hair length, we opt to use qualitative language to compute values for diversity measures.
However, some faces with extreme perceived ages are not appropriate for facial recognition technology.
With respect to age, we initially used the following qualitative measure prompts:
\begin{itemize}
\item 
$\bm x_{m_4}=$``\textit{A \{young, old\} person}.''
\end{itemize}

Trials that use qualitative age phrasing frequently produce images with extremely low or high ages, such as children or seniors.
Since these image samples do not typically reflect the sample population intended for facial recognition technology, we adapted prompts with quantitative language describing a specific population range that facial recognition technology is appropriate for:
\begin{itemize}
\item 
$\bm x_{m_4}=$``\textit{A person in their \{20, 50\}s}.''
\end{itemize}

\begin{figure*}[!ht]
\centering
\includegraphics[width=0.75\textwidth]{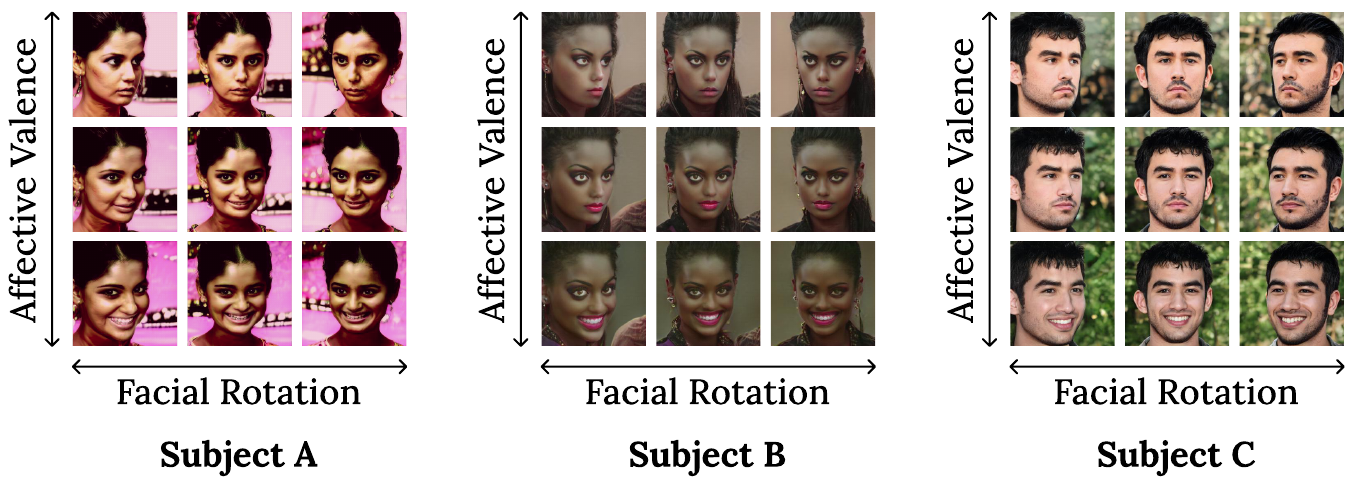}
\caption{Latent solutions are augmented with latent codes to increase intra-class variation. Each of $9$ data augmentations are visualized for $3$ subjects. Data augmentations result in unintended changes that are different for each subject.}
\label{fig:appendix_augmentations}
\end{figure*}

\subsection{Quality-diversity hyperparameters}

We adapt hyperparameters from the latent space illumination experiments described in~\cite{cma-mae}, to increase the degree of change at each iteration by increasing the learning rate $\eta$ and the initial branching step size $\sigma_g$.
These changes allow for more unmeasured diversity within the solution archive, since each solution deviates from its neighbors more in latent space.
Unmeasured diversity can be further increased with multiple trials of data generation by restarting with an empty passive elitist archive.

\paragraph{Shapes domain:} 
We run $1$ trial with the following hyperparameters.
We initialize the solution with $\bm\theta_0=\bm0$.
We define the number of iterations $N=7000$, learning rate $\eta=0.5$, population size $\lambda=32$, the initial branching step size $\sigma_g=0.5$, the annealing learning rate $\alpha=0.02$, and the minimum acceptable solution quality $\min_f=0$.

\paragraph{Facial recognition domain:}
We run $1$ trial to generate 15K solutions and $3$ trials for 50K solutions with the following hyperparameters.
We initialize the solution with $\bm\theta_0=\bm0$.
We define the number of iterations $N=6000$, learning rate $\eta=0.5$, population size $\lambda=32$, the initial branching step size $\sigma_g=0.5$, the annealing learning rate $\alpha=0.02$, and the minimum acceptable solution quality $\min_f=0$.

\section{Additional Results}
\subsection{Distributions of additional facial datasets}
In Table~\ref{tab:datasets-and-sampling-full}, we build on Table~\ref{tab:datasets-and-sampling} with evaluation benchmarks CALFW~\cite{calfw}, AgeDB~\cite{agedb}, CFPFP~\cite{cfpfp}, CPLFW~\cite{cplfw}.
We additionally include our datasets Rand15 and QD15 generated from random sampling and QDGS.

Notably, each evaluation benchmark contains large imbalances, particularly for skin tone.
For instance, AgeDB is nearly fully composed ($\approx99\%$) of faces with light or mixed skin tones.
Further, CALFW, CFPFP, and CPLFW each contain more than twice the amount of masculine to feminine faces.
We highlight that QDGS consistently produces more balanced proportions of labeled attributes corresponding to language measures (\textit{Skin Tone}, \textit{Age}, \textit{Gender}).

\subsection{Attribute conflation in data augmentations}
Generative models are typically paired with inverse image methods to identify latent codes that enable transformation of images toward desired attributes.
In our facial recognition experiments, we use latent codes for intra-class data augmentation of individual faces (Section~\ref{sec:data_augmentations}).

However, these codes are typically conflated with other semantic concepts and do not represent isolated features.
In other words, changing images along one dimension can have unpredictable effects on other image features.
Importantly, these changes do not always generalize across all faces and can be specific to a certain point in latent distribution.

We present a case study of these effects on $3$ subjects with each of the $9$ data augmentations, visualized in Fig.~\ref{fig:appendix_augmentations}.
The intended dimensions of augmentation are facial rotation and affective valence, but transformations typically result in other changes that are different for each individual:

\begin{itemize}
\item \textbf{Subject A:} Augmentations toward the bottom-right direction exhibit younger perceived age.
\item \textbf{Subject B:} Augmentations toward the top direction contain higher image brightness.
\item \textbf{Subject C:} Augmentations toward the right direction produce more dense facial hair.
\end{itemize}

Attribute conflations that are specific to certain faces can have potentially adverse implications.
If, for example, data augmentations systematically perturb identity-specific attributes more for dark-skinned faces than for light-skinned faces, facial recognition classifiers trained with this synthetic data may be more biased against dark-skinned users.
Due to this limitation, we believe the correct approach to implement latent code walks for synthetic data generation is to first inspect their impacts across different demographics.

This case study also broadly supports that StyleGAN2's latent space is biased along many measures, including those not identified in this paper.
This finding further highlights the need for diversity-aware sampling methods to identify and mitigate imbalances in synthetic training data.

\begin{figure*}[tbp]
\centering
\includegraphics[width=\textwidth]{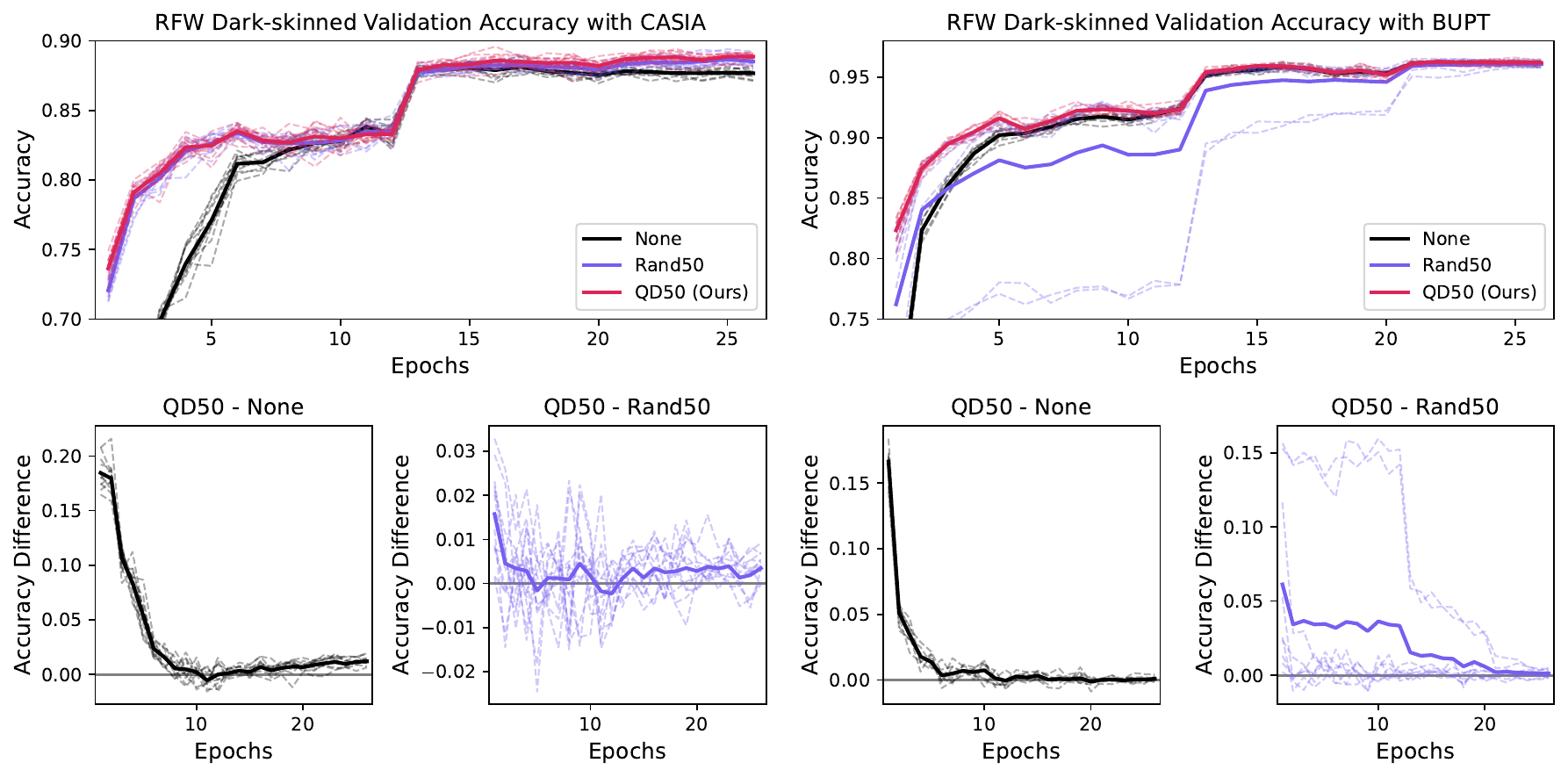}
\caption{\textbf{Top}: Training runs with CASIA and BUPT, separated by pretraining method. Average lines are darkened. \textbf{Bottom}: Differences between pretraining methods are typically larger in the beginning and taper off as they are trained on real datasets.}
\label{fig:appendix_training_runs}
\end{figure*}

\subsection{Effects of QDGS on training behavior}
While the final improvements in facial recognition performance are generally small, we nonetheless find that models pretrained with different datasets undergo different training behaviors.
In Fig.~\ref{fig:appendix_training_runs}, we show the RFW dark-skinned validation accuracy during training.
Across runs, we find that models pretrained with QD50 begin with higher accuracy than both models that are not pretrained and models that are pretrained with Rand50.
Intuitively, the difference between pretraining methods decreases with the number of epochs trained on real datasets.

\begin{table}[tbp]
  \small
  \centering
  \setlength{\tabcolsep}{4pt}
  \begin{tabular}{l c l l}
    \toprule
    \textbf{Dataset} &  &
    \multicolumn{2}{c}{\textbf{Approximate run time}} \\
    \toprule
    \multicolumn{1}{l}{Data Generation}\\
    \cmidrule(r){1-1}
    Shapes && $\phantom{0}1$ hour & $\phantom{0000}-$\\
    Facial Recognition (15K solutions) && $\phantom{0}4$ hours & $\phantom{0000}-$\\
    Facial Recognition (50K solutions) && $17$ hours & $30$ minutes\\
    \toprule
    \multicolumn{1}{l}{Classifier Training}\\
    \cmidrule(r){1-1}
    Shapes & & $\phantom{0000}-$ & $\phantom{0}5$ minutes\\
    Facial Recognition (15K solutions) && $\phantom{0}5$ hours & $30$ minutes\\
    Facial Recognition (50K solutions) && $13$ hours & $\phantom{0000}-$\\
    Facial Recognition (CASIA) && $16$ hours & $\phantom{0000}-$\\
    Facial Recognition (BUPT) && $25$ hours & $30$ minutes\\
    \bottomrule
  \end{tabular}
  \caption{Approximate run times on a AMD EPYC 7513 CPU and NVIDIA Tesla A100 GPU, without parallelization.}
  \label{tab:appendix_runtimes}
\end{table}

\section{Reproducibility}

\subsection{Source code}
The source code to use QDGS for the shape and facial recognition domains is provided at the following link:
\href{https://github.com/Cylumn/qd-generative-sampling}{\texttt{github.com/Cylumn/qd-generative-sampling}}.
The \texttt{README.md} document contains information to install the required dependencies and to run the data generation, training, and evaluation scripts.

\subsection{Compute}
We run experiments on a high-performance computing cluster.
We use compute nodes equipped with distinct resource combinations: Intel Xeon E5-2640 v4 CPUs and NVIDIA Tesla P100 GPUs, Intel Xeon Gold 6130 CPUs and NVIDIA Tesla V100 GPUs, and AMD EPYC 7513 CPUs and NVIDIA Tesla A100 GPUs.

\subsection{Run times}
We parallelize data generation for experiments that run multiple trials of QDGS, which can greatly improve efficiency.
The approximate run times without parallelization for data generation and classifier training are reported in Table~\ref{tab:appendix_runtimes}.}\fi
\end{document}